\documentclass[a4paper,fleqn,usenatbib]{mnras}

\usepackage{newtxtext,newtxmath}

\usepackage[T1]{fontenc}
\usepackage[utf8]{inputenc}
\usepackage{ae,aecompl}

\usepackage{graphicx}	
\usepackage{xcolor}
\usepackage{amsmath}	
\usepackage{amssymb}	
\usepackage{amsbsy}
\usepackage{multicol}   
\usepackage{bm}		    
\usepackage{pdflscape}	
\usepackage{subfigure}
\usepackage{hyperref}
\usepackage{latexsym}
\usepackage{marvosym}
\usepackage{enumerate}
\usepackage{textcomp}
\usepackage{float}

\renewcommand{\vec}[1]{{\boldsymbol #1}}

\title[High mode magnetohydrodynamic waves in a rotating jet]{High mode magnetohydrodynamic waves propagation in a twisted rotating jet emerging from a filament eruption}

\author[I.~Zhelyazkov and R.~Chandra]{
Ivan~Zhelyazkov$^{1}$\thanks{E-mail: izh@phys.uni-sofia.bg}
and Ramesh~Chandra$^{2}$
\\
$^{1}$Faculty of Physics, Sofia University, 1164 Sofia, Bulgaria\\
$^{2}$Department of Physics, DSB Campus, Kumaun University, Nainital 263002, India
}

\date{Accepted XXX. Received YYY; in original form ZZZ}

\pubyear{2018}

\begin{document}
\label{firstpage}
\pagerange{\pageref{firstpage}--\pageref{lastpage}}
\maketitle

\begin{abstract}
We study the conditions under which high mode magnetohydrodynamic (MHD) waves propagating on a rotating jet emerging from the filament eruption on 2013 April 10--11 can became unstable against the Kelvin--Helmholtz instability (KHI).  The evolution of jet indicates the blob like structure at its boundary which could be one of the observable features of the KHI development.  We model the jet as a twisted rotating axially moving magnetic flux tube and explore the propagation characteristics of the running MHD modes on the basis of dispersion relations derived in the framework of the ideal magnetohydrodynamics.  It is established that unstable MHD waves with wavelengths in the range of $12$--$15$~Mm and instability developing times from $1.5$ to $2.6$~min can be detected at the excitation of high mode MHD waves.  The magnitude of the azimuthal mode number $m$ crucially depends upon the twist of the internal magnetic field.  It is found that at slightly twisted magnetic flux tube the appropriate azimuthal mode number is $m = 16$ while in the case of a moderately twisted flux tube it is equal to $18$.
\end{abstract}




\section{Introduction}
\label{sec:intro}

Rotating, tornado-like jets are among the most spectacular events in the solar atmosphere.  They have been studied for almost a century with ground-based and borne-craft instruments primarily for their strong impact on the space whether on the Earth -- for a review of both observation and modeling see, e.g., \citet{Wedemeyer2012,Zhelyazkov2018} and references therein.  We are not going to comment on the large number of publications discussing the prominences tornadoes \citep[e.g.,][and references therein]{Wedemeyer2013a,Wedemeyer2013b,Wedemeyer2014,Su2014,Levens2016}, but will focus ourselves on small-scale active phenomena on the Sun such as rotating magnetically twisted jets, be them macrospicules \citep{Pike1998,Kamio2010,Curdt2011,Bennett2015,Kiss2017,Kiss2018}, Type II spicules \citep{DePontieu2012,Martinez-Sykora2013}, X-ray jets \citep{Moore2013}, or coronal hole EUV jets \citep{Nistico2009,Liu2009,Nistico2010,Shen2011,Chen2012,Hong2013,Young2014a,Young2014b,Moore2015}.  A coronal rotating EUV jet can emerge form a swirling flare as it has been observed by \citet{Zhang2014}.  A helically twisted plasma jet can also be formed during a confined filament eruption \citep{Filippov2015} within a null-point topology known as an inverted \textsf{Y} magnetic field configuration.  Namely that jet will be the target of our study.

The physical parameters of rotating jets can be quite different depending on the jet's nature.  Electron number densities vary (in descending order) from $10^{11}$ through $10^9$ to $10^8$~cm$^{-3}$, while the electron temperatures lie between $10^4$ and $(3$--$4)\times 10^6$~K.  The flow velocities are in the range of $20$ to $400$~km\,s$^{-1}$ and the rotating ones can be from $10$ to $200$~km\,s$^{-1}$.  The jet's width also varies depending on jet's nature, but sizes of a few to $30$~megameters are typical for observed eruptive events.  The heights/lengths can be rather different too, from a few Mm to $200$ or more Mm.  Jets's lifetimes lie in the wide range from a few minutes (or less) to, say, $30$~min.  An extensive study of the origin and the physical parameters especially for coronal hole jets the reader can find in \citet{Raouafi2016}.

Solar jets that are magnetically structured configurations support the propagation of various types of MHD waves (fast and slow magnetoacoustic waves, Alfv\'en waves) which must be considered as normal MHD modes running in a given jet.  The most natural model of a solar jet is a moving with velocity $\vec{U}$ cylindrically shaped magnetic flux tube with an electron number density $n_\mathrm{i}$, surrounded by immobile/moving plasma with electron number density $n_\mathrm{e}$.  Generally, one assumes that the magnetic field inside the jet, $\vec{B}_\mathrm{i}$, and the one of the environment, $\vec{B}_\mathrm{e}$, are different in magnitude and topology.  The subscript labels `i' and `e' stamp for \emph{interior\/} and \emph{exterior}, respectively.  Each jet can potentially become unstable against any instability.  Such an instability is the Kelvin--Helmholtz instability (KHI) that arises at the interface of two incompressible plasmas moving with different velocities embedded in a constant magnetic field if the thin velocity shear around the interface exceeds some critical value \citep{Chandrasekhar1961}.  In cylindrical geometry, the KHI in its nonlinear stage can develop a series of KH vortices at a thin shell near the tube boundary.  KHI plays an important role because it can trigger the plasma wave turbulence which along with the micro/nano magnetic reconnection, yielding microflares/nanoflares, is considered as one of the main heating mechanisms of the solar corona \citep{Cranmer2015}.

Over past few decades KHI was studied in various jets's magnetic field configurations (slab or cylindrical geometry) in the solar atmosphere, solar wind, and Earth magnetosphere.  The most interesting for our study are the cases in which the KHI was observationally detected and appropriately modeled.  A prime example for that is the KHI in coronal mass ejections, firstly observed by \citet{Foullon2011} and \cite{Ofman2011} and later on modeled in a series of articles \citep{Foullon2013,Mostl2013,Nykyri2013,Zhelyazkov2015}.  A similar KH vortex pattern, showing moving blobs along the boundary of a rotating solar coronal jet arising close to the north pole of the Sun \citep{Chen2012}, was modeled \citep{Zhelyazkov2018} and it has been shown that these blobs are the manifestation of the KHI of a high azimuthal mode number (${=}12$) MHD wave propagating along the jet.  From the graphic plots picturing the dependencies of normalized wave phase velocity and wave growth rate on the normalized wavenumber was deduced that at the unstable wavelength of $12$~Mm the instability evolution/growth time is $4.7$~min, while at $\lambda_\mathrm{KH} = 15$~Mm it is $1.9$~min, both values in good agreement with the analysis of the observational data.

The numerical solutions to the wave dispersion relation governing the propagation of the excited MHD mode depend on a couple of input parameters.  One of them is the density contrast between the environment and the jet itself defined as the ratio $n_\mathrm{e}/n_\mathrm{i}$, which at given electron temperatures $T_\mathrm{i}$ and $T_\mathrm{e}$ and background magnetic filed $B_\mathrm{e}$ determines the plasma betas in both media.  It turns out that plasma betas are crucial in the KHI modeling -- if the betas are close or bigger than $1$, both media can be treated as quasi-incompressible/incompressible plasmas \citep{Zank1993}.  If, however, any beta is less/much less than $1$, the corresponding medium should be considered as a cool plasma.  Another input parameter, which can control the width of the wavenumbers/wavelengths ranges in which KHI occurs, is the jet's magnetic field twist.  It this study, we investigate how the choice of the internal magnetic field twist affects the MHD mode and instability characteristics in the rotating jet arising during the filament eruption on 2013 April 10--11 and observationally explored by \citet{Filippov2015}.  The organization of the article is as follows: in the next section we present and discuss the observational data concerning jet \#2 in \citet{Filippov2015}.  Section 3 is devoted to the jet geometry, magnetic and velocity fields topologies, basic physical parameters of the jet and its environment, as well as the equations governing the MHD mode propagation in the system.  Section 4 deals with the results of the numerical solutions to the appropriate dispersion relation.  In the last section, we summarize the main results obtained in this study.

\section{Observations}
\label{sec:observations}
The observational study of jet events of 2013 April 10--11 was done by \citet{Filippov2015}.
In their study they report three jets originated from the active region NOAA 11715.  The active region
was located on the west limb during that period.  These authors found that the confined eruption of the filament having null-point topology results in the formation of observed jets.  Out of these three jets we modeled KHI for the second jet.  For modeling the KHI, we need different physical parameters such as: temperature and the density inside and outside of the jet.  Therefore, we have calculated these parameters using techniques proposed by \citet{Aschwanden2013}.  This techniques needs the data of six  Atmospheric Imaging Assembly (AIA, \citealp{Lemen2012}) onboard \emph{Solar Dynamics Observatory\/} (\emph{SDO}, \citealp{Pesnell2012}) satellite EUV channels, i.e., 94, 131, 171, 193, 211 and 335~\AA.  The estimated values of temperature inside and outside are 2.0~MK and 2.14~MK, respectively.  In addition to this, the calculated values of number densities inside and outside of the jet are $4.65 \times 10^9$~cm$^{-3}$ and $4.02 \times 10^9$~cm$^{-3}$, respectively.  The jet's width is estimated to be ${\approx}30$~Mm, its height $180$~Mm, and life time $30$~min.  The basic physical parameters of two media with their error estimates and corresponding plasma betas, calculated at background magnetic field of $5$~G are given in Table~\ref{tab:parameter}.  We note, that the plasma beta for each medium was calculated from the expression $(6/5)c_\mathrm{s}^2/v_\mathrm{A}^2$, where $c_\mathrm{s} = (\gamma k_\mathrm{B}T_\mathrm{e}/m_\mathrm{i})^{1/2}$ is the sound speed (with $\gamma$ being the adiabatic index equal to $5/3$, $k_\mathrm{B}$ the Boltzmann's constant, $T_\mathrm{e}$ the electron temperature, and $m_\mathrm{i}$ the ion/proton mass), and $v_\mathrm{A} = B/(\mu n_\mathrm{i}m_\mathrm{i})^{1/2}$ is the Alfv\'en speed, in which $B$ is the full magnetic field (${=}(B_\phi^2 + B_z^2)^{1/2}$), $\mu$ is the vacuum magnetic permeability, and $n_\mathrm{i}$ is the ion/proton number density.  Aforementioned formula for plasma beta follows from its definition as the ratio of thermal to the magnetic pressure.

\begin{table}
\caption{Different jet's and its environment physical parameters derived from \emph{SDO}/AIA data at $B_\mathrm{e} = 5$~G.}
\label{tab:parameter}
\vspace*{2mm}

\centering

\begin{tabular}{cccc}
\hline
Medium & Temperature & Electron density & Plasma beta \\
       & (MK)        &  (${\times}10^9$ cm$^{-3}$) & \\
\hline
Jet & $2.00 \pm 0.02$ & $4.65 \pm 0.06$ & $0.596$  \\
Env & $2.14 \pm 0.01$ & $4.02 \pm 0.04$ & $1.196$  \\
\hline
\end{tabular}
\end{table}
\begin{figure}
\centering
    \includegraphics[width=7.5cm]{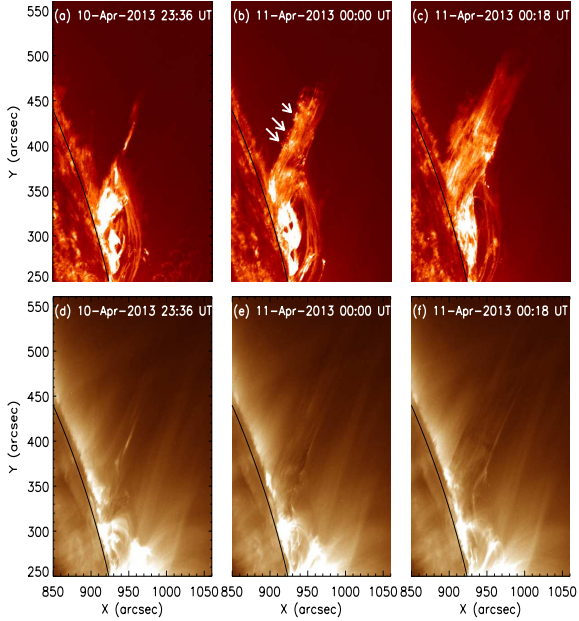}
  \caption{Evolution of the jet in AIA 304~\AA{}(top) and 193~\AA{} (bottom).}
   \label{fig:fig1}
\end{figure}

The evolution of the jet in AIA 304 and 193~\AA\ is displayed in Fig.~\ref{fig:fig1}.  According to the study of \citet{Filippov2015}, we have also found that the rotation is in anti-clockwise direction.  During the evolution of the jet, we have observed the vortex type structures in the eastern side of the jet.  These vortex type structures are indicated by the arrows and can be considered as an evidence of the KHI.  As shown in the article of \citet{Filippov2015}, the jet location in the \emph{STEREO A\/} \citep{Kaiser2008} data was close to the location of a coronal hole.  The coronal holes are the locations, where the magnetic field line are open. These open nearby magnetic structures provides the opportunity for an easy escape of the jets.

\section{Geometry, magnetic field, and governing equations}
\label{sec:geometry}
We model the jet, as already have been said in the Introduction section, as a moving with velocity $\vec{U}$ twisted magnetic flux tube of radius $a$ and homogeneous electron number density $n_\mathrm{i}$, or in other words, homogeneous plasma density $\rho_\mathrm{i}$.  That tube is surrounded by a plasma with constant density $\rho_\mathrm{e}$ embedded in a homogeneous magnetic field $\vec{B}_\mathrm{e}$ which, in cylindrical coordinates ($r, \phi, z$) has only a $z$ component, that is,  $\vec{B}_\mathrm{e} = (0, 0, B_\mathrm{e})$.  By contrast, both the magnetic field and flow velocity inside the magnetic flux tube are assumed to be twisted and presented by the vectors $\vec{B}_\mathrm{i} = \left( 0, B_{\mathrm{i}\phi}(r), B_{\mathrm{i}z} \right)$ and $\vec{U} = \left( 0, U_{\phi}(r), U_{z} \right)$, respectively.  We note that the $z$ components of the twisted magnetic field, $B_{\mathrm{i}z}$, and twisted jet velocity, $U_z$, are constant.  Under these circumstances, the pressure balance condition inside the jet, derived from the integration of the momentum equation for the equilibrium variables, yields the following radial profile of the total pressure \citep{Zaqarashvili2015}:
\begin{equation}
\label{eq:pressurebalance}
    p_\mathrm{t}(r) =  p_\mathrm{t}(0) - \frac{1}{\mu}\int\displaylimits_0^r \frac{B_{\mathrm{i}\phi}^2(s)}{s}\mathrm{d}s + \rho_\mathrm{i} \int\displaylimits_0^r \frac{U_\phi^2(s)}{s} \mathrm{d}s,
\end{equation}
where $\mu$ is the plasma permeability and $p_\mathrm{t}(0)$ is the total (thermal $+$ magnetic) pressure.

For simplicity we assume that the two azimuthal components of the internal magnetic field and flow velocity are linear functions of the radial position, $s$, that is, $B_{\mathrm{i}\phi}(s) = As$ and $U_\phi(s) = \Omega s$, respectively, where $A$ and $\Omega$ are constants.  Thus, the rotating jet velocity at the boundary, $U_\phi (a) \equiv U_\phi$, determined from observations, in rigid rotation case, can be expressed through the jet angular velocity, $\Omega$, and tube radius, $a$, through the relation $U_\phi = \Omega a$.  In a similar way, we can denote the magnetic field azimuthal component at the tube boundary as $B_{\mathrm{i}\phi}(a) \equiv B_{\phi} = Aa$.  Since we will treat both media as incompressible or cool media (depending on the case), then the total pressure balance equation for the jet--environment system, after performing the integration in Eq.~(\ref{eq:pressurebalance}) from zero to the tube radius $a$, takes the form
\begin{equation}
\label{eq:pbeq}
    p_\mathrm{i} - \frac{1}{2}\rho_\mathrm{i}U_\phi^2 + \frac{B_{\mathrm{i}z}^2}{2\mu}\left( 1 + \varepsilon_1^2 \right) = p_\mathrm{e} + \frac{B_\mathrm{e}^2}{2\mu},
\end{equation}
where $\varepsilon_1 \equiv B_{\phi}/B_{\mathrm{i}z} = Aa/B_{\mathrm{i}z}$ is the magnetic field twist parameter.  Similarly we introduce $\varepsilon_2 \equiv U_\phi/U_z = \Omega a/U_z$, that characterized the jet velocity twist.  Here, $p$ denotes the thermal/plasma pressure.  We note that in our case $\varepsilon_2$ is defined by observationally measured rotational and axial velocities while, at least for now, $\varepsilon_1$ is a free parameter, that is, on using Eq.~(\ref{eq:pbeq}) we must specify its value.

The basic jet--environment physical parameters, recall, are: electron number densities $n_\mathrm{i} = 4.65 \times 10^9$~cm$^{-3}$, $n_\mathrm{e} = 4.02 \times 10^9$~cm$^{-3}$ and electron temperatures $T_\mathrm{i} = 2.0$~MK, $T_\mathrm{e} = 2.14$~MK, respectively.  The axial velocity of the jet is $U_z = 100$~km\,s$^{-1}$, the azimuthal one is $U_\phi = 180$~km\,s$^{-1}$, the jet width is $\Delta \ell \approx 30$~Mm, while its height is $H = 180$~Mm.  With the aforementioned electron number densities, the density contrast, which we denote with $\eta$, has a magnitude of $0.864$.  Assuming that the background magnetic field is $B_\mathrm{e} = 5$~G, depending on the value of the internal magnetic field twist parameter  $\varepsilon_1$, on using the total pressure balance equation, we can have different Alfv\'en speeds, as well as different values of the ratio of the external to the internal axial magnetic field components, $b = B_\mathrm{e}/B_{\mathrm{i}z}$.  This $b$ is also an input parameter in the numerical calculations.  For convenience, we define the Alfv\'en speed in given medium as $v_\mathrm{A} = B_z/\!\sqrt{\mu \rho}$.  The main goal of our study, as we already said in the Introduction section, is to see how the choice of the internal magnetic field twist parameter, $\varepsilon_1$, will modify KHI characteristics, notably the frequency growth rate, instability development/growth time, and the phase velocity of the unstable mode.  In the following, we will consider two cases, namely the case of a slightly twisted moving flux tube with $\varepsilon_1 = 0.005$, and the case of an intermediately twisted tube with a twist of $0.1$ which is reasonable for our jet -- that value was obtained from the inclination of the jet's treads in the event on 2013 April 10 detected by \emph{SDO}/AIA as the inclination of jet's threads yields the relationship between axial and azimuthal magnetic field components, i.e., $\varepsilon_1$.  With these two values of $\varepsilon_1$, the corresponding Alfv\'en speeds inside the jet are equal to $235.53$ and $234.36$~km\,s$^{-1}$, respectively, while the ratio of the axial magnetic field components, $b$, possesses magnitudes correspondingly of $0.679$ and $0.682$.

As seen from Table~\ref{tab:parameter}, jet's plasma beta is less than one and it is more realistic the jet to be treated as a cool plasma while its environment with $\beta_\mathrm{e} \cong 1.2$ can be considered as an incompressible medium.
Dispersion relation of high MHD modes propagating in moving rotating cylindrical jets in the limit of incompressible plasmas was derived by \citet{Zaqarashvili2015}, but it is inapplicable in our case of cool jet surrounded by incompressible magnetized plasma.  Hence, one needs to derive a new wave dispersion equation being adequate to the present jet--environment system.  That derivation is exposed in the next subsection.

\subsection{MHD wave dispersion equation in a cool jet plasma--incompressible environment system}
\label{subsec:disp_eq}

Excited MHD waves propagate in axial direction along the rotating twisted magnetic flux tube that implies a wavevector
$\vec{k} = (0, 0, k_z)$.  For a zero beta plasma, we consider the small perturbations from equilibrium in the form
\[
    \vec{B} = \vec{B}_0 + \vec{B}_1, \quad \rho = \rho_0 + \rho_1, \quad \vec{v} = \vec{v}_0 + \vec{v}_1,
\]
where the subscript $0$ denotes the equilibrium values of magnetic and velocity fields, while the subscript $1$ denotes their perturbations.  The plasma motion is described by the linearized ideal MHD equations which have the form
\begin{equation}
\label{eq:cont}
    \frac{\partial \rho_1}{\partial t} = -\nabla \cdot (\rho_0 \vec{v}_1 + \rho_1 \vec{v}_0) = 0,
\end{equation}
\begin{equation}
\label{eq:momentum}
    \rho_0 \left( \frac{\partial}{\partial t} + \vec{v}_0 \cdot \nabla \right)\vec{v}_1 = \frac{1}{\mu}(\nabla \times \vec{B}_1) \times \vec{B}_0,
\end{equation}
\begin{equation}
\label{eq:faraday}
    \frac{\partial \vec{B}_1}{\partial t} = \nabla \times (\vec{v}_0 \times \vec{B}_1) + \nabla \times (\vec{v}_1 \times \vec{B}_0),
\end{equation}
and the constraint
\begin{equation}
\label{eq:divB}
    \nabla \cdot \vec{B}_1 = 0.
\end{equation}
Recall that for cold plasmas the total pressure reduces to the magnetic pressure only, that is $p_\mathrm{t} = p_\mathrm{m}$, the $z$ component of the velocity perturbation is zero, i.e., $\vec{v}_1 = (v_{1r}, v_{1\phi}, 0)$, while $\vec{B}_1 = (B_{1r}, B_{1\phi}, B_{1z})$.  Equation~(\ref{eq:cont}), which defines the density perturbation, is not used below because we are studying the propagation and stability of Alfv\'en-wave like perturbations of the fluid velocity and magnetic field.  Bearing in mind that here $\vec{B}_0 \equiv \vec{B}_\mathrm{i} = \left( 0, B_{\mathrm{i}\phi}(r), B_{\mathrm{i}z} \right)$ and $\vec{v}_0 \equiv \vec{U} = \left( 0, U_\phi(r), U_z \right)$, the above set of vector equations reduces to the following $6$ differential equations:
\begin{eqnarray}
\label{eq:v1r}
    \left( \frac{\partial}{\partial t} + U_\phi \frac{1}{r}\frac{\partial}{\partial \phi} + U_z \frac{\partial}{\partial z} \right)v_{1r} - 2\frac{U_\phi}{r}v_{1\phi} - \frac{1}{\mu \rho_\mathrm{i}} \nonumber \\
    \nonumber \\
    {}\times \left[ B_{\mathrm{i}\phi}\frac{1}{r}\frac{\partial}{\partial \phi} + B_{\mathrm{i}z}\frac{\partial}{\partial z} \right]B_{1r} + 2 \frac{B_{\mathrm{i}\phi}}{\mu \rho_\mathrm{i}r}B_{1\phi} = -\frac{1}{\rho_\mathrm{i}} \frac{\partial p_{\mathrm{m}1}}{\partial r},
\end{eqnarray}
\begin{eqnarray}
\label{eq:v1phi}
    \left( \frac{\partial}{\partial t} + U_\phi \frac{1}{r}\frac{\partial}{\partial \phi} + U_z \frac{\partial}{\partial z} \right)v_{1\phi} + \frac{1}{r}\frac{\partial(rU_\phi)}{\partial r}v_{1r} \nonumber \\
    \nonumber \\
    {}- \frac{1}{\mu \rho_\mathrm{i}} \left[ B_{\mathrm{i}\phi}\frac{1}{r}\frac{\partial}{\partial \phi}
    + B_{\mathrm{i}z}\frac{\partial}{\partial z} \right]B_{1\phi} - \frac{1}{\mu \rho_\mathrm{i}r} \frac{\partial(rB_{\mathrm{i}\phi})}{\partial r}B_{1r} \nonumber \\
    \nonumber \\
    {}= -\frac{1}{\rho_\mathrm{i}}\frac{1}{r} \frac{\partial p_{\mathrm{m}1}}{\partial \phi},
\end{eqnarray}
\begin{equation}
\label{eq:v1z}
    \frac{1}{\mu \rho_\mathrm{i}}\left[ B_{\mathrm{i}\phi}\frac{1}{r}\frac{\partial}{\partial \phi} + B_{\mathrm{i}z}\frac{\partial}{\partial z} \right]B_{1z} = \frac{1}{\rho_\mathrm{i}}\frac{\partial p_{\mathrm{m}1}}{\partial z},
\end{equation}
\begin{eqnarray}
\label{eq:B1r}
    \left( \frac{\partial}{\partial t} + U_\phi \frac{1}{r}\frac{\partial}{\partial \phi} + U_z \frac{\partial}{\partial z} \right)B_{1r} - \left[ B_{\mathrm{i}\phi}\frac{1}{r}\frac{\partial}{\partial \phi} + B_{\mathrm{i}z}\frac{\partial}{\partial z} \right]v_{1r} \nonumber \\
    \nonumber \\
    {}= 0,
\end{eqnarray}
\begin{eqnarray}
\label{eq:B1phi}
    \left( \frac{\partial}{\partial t} + U_\phi \frac{1}{r}\frac{\partial}{\partial \phi} + U_z \frac{\partial}{\partial z} \right)B_{1\phi} - r\frac{\partial}{\partial r}\left(\frac{U_\phi}{r}\right)B_{1r} \nonumber \\
    \nonumber \\
    {}- \frac{1}{\mu \rho_\mathrm{i}} \left[ B_{\mathrm{i}\phi}\frac{1}{r}\frac{\partial}{\partial \phi}
    {}+ B_{\mathrm{i}z}\frac{\partial}{\partial z} \right]v_{1\phi} + r\frac{\partial}{\partial r}\left( \frac{B_{\mathrm{i}\phi}}{r} \right)v_{1r} = 0,
\end{eqnarray}
\begin{equation}
\label{eq:nablaB}
    \frac{\partial}{\partial r}B_{1r} + \frac{1}{r}B_{1r} + \frac{1}{r} \frac{\partial}{\partial \phi}B_{1\phi} + \frac{\partial}{\partial z}B_{1z} = 0.
\end{equation}

To investigate the stability of the system, Eqs.~(\ref{eq:v1r})--(\ref{eq:nablaB}) are Fourier transformed, assuming that all perturbations, in cylindrical coordinates, have the form
\begin{equation}
\label{eq:fourier}
    g(r,\phi,z,t) = g(r)\exp[\mathrm{i}(-\omega t + m \phi + k_z z)],
\end{equation}
where $g$ represents any quantities $\vec{v}_1$, $p_{\mathrm{m}1}$, and $\vec{B}_1$; $\omega$ is the angular wave frequency, $m$ is the azimuthal mode number, and $k_z$ is the axial wavenumber.  Taking into account that according to our specific choice of uniform twists of the internal magnetic field and the jet velocity, where $B_{\mathrm{i}\phi}(r) = Ar$ and $U_{\phi}(r) = \Omega r$, if we use Eq.~(\ref{eq:fourier}) in Eqs.\ (\ref{eq:v1r})--(\ref{eq:nablaB}), we obtain the following set of equations for the components of the fluid velocity, magnetic pressure, and magnetic field perturbations:
\begin{eqnarray}
\label{eq:v1rnew}
    -\mathrm{i}\sigma v_{1r} - 2\Omega v_{1\phi} - \mathrm{i}\frac{1}{\mu \rho_\mathrm{i}}\left( mA + k_z B_{\mathrm{i}z} \right)B_{1r} + 2\frac{A}{\mu \rho_\mathrm{i}}B_{1\phi} \nonumber \\
    \nonumber \\
    {}= -\frac{1}{\rho_\mathrm{i}} \frac{\mathrm{d}p_{\mathrm{m}1}}{\mathrm{d}r},
\end{eqnarray}
\begin{eqnarray}
\label{eq:v1phinew}
    -\mathrm{i}\sigma v_{1\phi} + 2\Omega v_{1r} - \mathrm{i}\frac{1}{\mu \rho_\mathrm{i}}\left( mA + k_z B_{\mathrm{i}z} \right)B_{1\phi} - 2\frac{A}{\mu \rho_\mathrm{i}}B_{1r} \nonumber \\
    \nonumber \\
    {}= -\mathrm{i}\frac{m}{r}\frac{1}{\rho_\mathrm{i}} p_{\mathrm{m}1},
\end{eqnarray}
\begin{equation}
\label{eq:v1znew}
    \frac{1}{\mu \rho_\mathrm{i}}\left( mA + k_z B_{\mathrm{i}z} \right)B_{1z} = \frac{1}{\rho_\mathrm{i}}k_z p_{\mathrm{m}1},
\end{equation}
\begin{equation}
\label{eq:B1rnew}
    \sigma B_{1r} - \left( mA + k_z B_{\mathrm{i}z} \right)v_{1r} = 0,
\end{equation}
\begin{equation}
\label{eq:B1phinew}
    \sigma B_{1\phi} - \left( mA + k_z B_{\mathrm{i}z} \right)v_{1\phi} = 0,
\end{equation}
\begin{equation}
\label{eq:nablaBnew}
    \left(\frac{\mathrm{d}}{\mathrm{d} r} + \frac{1}{r} \right)B_{1r} + \mathrm{i}\frac{m}{r}B_{1\phi} + \mathrm{i}k_z B_{1z} = 0.
\end{equation}
where $p_{1\mathrm{m}} = B_{\mathrm{i}z}B_{1z}/\mu$ is the magnetic pressure perturbation and $\sigma = \omega - m\Omega - k_z U_z$ is the Doppler shifted frequency.

By defining
\begin{equation}
\label{eq:omegaAi}
    \omega_\mathrm{Ai} = \frac{1}{\sqrt{\mu \rho_\mathrm{i}}}\left( mA + k_z B_{\mathrm{i}z} \right),
\end{equation}
which is the local Alfv\'en frequency, Eqs.~(\ref{eq:v1rnew}) and (\ref{eq:v1phinew}) take the form
\[
    -\mathrm{i}\sigma v_{1r} - 2\Omega v_{1\phi} - \mathrm{i}\frac{1}{\sqrt{\mu \rho_\mathrm{i}}}\omega_\mathrm{Ai}B_{1r} + \frac{2A}{\mu \rho_\mathrm{i}}B_{1\phi} + \frac{1}{\rho_\mathrm{i}} \frac{\mathrm{d}}{\mathrm{d}r}p_{\mathrm{m}1} = 0,
\]
\[
    -\mathrm{i}\sigma v_{1\phi} + 2\Omega v_{1r} - \mathrm{i}\frac{1}{\sqrt{\mu \rho_\mathrm{i}}}\omega_\mathrm{Ai}B_{1\phi} - \frac{2A}{\mu \rho_\mathrm{i}}B_{1r} + \mathrm{i}\frac{1}{\rho_\mathrm{i}} \frac{m}{r}p_{\mathrm{m}1} = 0.
\]
By replacing $B_{1r}$ and $B_{1\phi}$, expressed from Eqs.~(\ref{eq:B1rnew}) and (\ref{eq:B1phinew}), into the above equations, after some algebra one obtains
\begin{equation}
\label{eq:help1}
    -\mathrm{i}v_{1r} - Zv_{1\phi} + \frac{\sigma}{\sigma^2 - \omega_\mathrm{Ai}^2}\frac{1}{\rho_\mathrm{i}} \frac{\mathrm{d}}{\mathrm{d}r}p_{\mathrm{m}1} = 0,
\end{equation}
\begin{equation}
\label{eq:help2}
    -\mathrm{i}v_{1\phi} + Zv_{1r} + \mathrm{i}\frac{\sigma}{\sigma^2 - \omega_\mathrm{Ai}^2}\frac{1}{\rho_\mathrm{i}} \frac{m}{r}p_{\mathrm{m}1} = 0,
\end{equation}
where
\begin{equation}
\label{eq:Z}
    Z = 2\frac{\sigma \Omega + A\omega_\mathrm{Ai}/\sqrt{\mu \rho_\mathrm{i}}}{\sigma^2 - \omega_\mathrm{Ai}^2}.
\end{equation}
From Eq.~(\ref{eq:help1}) we obtain that
\[
    v_{1r} = -\mathrm{i}\frac{\sigma}{\sigma^2 - \omega_\mathrm{Ai}^2}\frac{1}{\rho_\mathrm{i}} \frac{\mathrm{d}}{\mathrm{d}r}p_{\mathrm{m}1} + \mathrm{i}Zv_{1\phi}.
\]
Substituting this expression into Eq.~(\ref{eq:help2}), we get an expression of $v_{1\phi}$ in terms of the magnetic pressure perturbation $p_{\mathrm{m}1}$:
\begin{equation}
\label{eq:v1phifinal}
    v_{1\phi} = \frac{1}{Y}\frac{\sigma}{\sigma^2 - \omega_\mathrm{Ai}^2}\frac{1}{\rho_\mathrm{i}}\left( \frac{m}{r} - Z\frac{\mathrm{d}}{\mathrm{d}r} \right)p_{\mathrm{m}1},
\end{equation}
where
\begin{equation}
\label{eq:Y}
    Y = 1 - Z^2.
\end{equation}
Going back a few lines above to the expression of $v_{1r}$ via $\mathrm{d}p_{\mathrm{m}1}/\mathrm{d}r$ and $v_{1\phi}$, after plugging there expression~(\ref{eq:v1phifinal}), we obtain an updated formula for $v_{1r}$, notably
\begin{equation}
\label{eq:v1rfinal}
    v_{1r} = -\mathrm{i}\frac{1}{Y}\frac{\sigma}{\sigma^2 - \omega_\mathrm{Ai}^2}\frac{1}{\rho_\mathrm{i}}\left( \frac{\mathrm{d}}{\mathrm{d}r} - Z\frac{m}{r} \right)p_{\mathrm{m}1}.
\end{equation}
From Eq.~(\ref{eq:v1znew}), after multiplying the two sides by $k_z$, we obtain that
\[
    \frac{1}{\rho_\mathrm{i}}k_z^2 p_{\mathrm{m}1} = \frac{1}{\sqrt{\mu \rho_\mathrm{i}}}\omega_\mathrm{Ai}k_z B_{1z},
\]
which yields
\[
    k_z B_{1z} = \sqrt{\frac{\mu}{\rho_\mathrm{i}}}\frac{1}{\omega_\mathrm{Ai}}k_z^2 p_{\mathrm{m}1}.
\]
After multiplying Eq.~(\ref{eq:nablaBnew}) by $-\mathrm{i}$, we have
\[
    -\mathrm{i}\left( \frac{\mathrm{d}}{\mathrm{d}r} + \frac{1}{r} \right)B_{1r} + \frac{m}{r}B_{1\phi} + k_z B_{1z} = 0.
\]
Here, we replace $B_{1r}$ and $B_{1\phi}$, obtained from Eqs.~(\ref{eq:B1rnew}) and (\ref{eq:B1phinew}), but using the new presentations of $v_{1r}$ and $v_{1\phi}$ via Eqs.~(\ref{eq:v1rfinal}) and (\ref{eq:v1phifinal}) in the above equation along with inserting in it the expression of $k_z B_{1z}$, to get after some algebra the following second order ordinary differential equation:
\begin{equation}
\label{eq:depm1}
    \left[ \frac{\mathrm{d^2}}{\mathrm{d}r^2} + \frac{1}{r}\frac{\mathrm{d}}{\mathrm{d}r} - \left( \frac{m^2}{r^2} + \kappa^2 \right) \right]p_{\mathrm{m}1} = 0,
\end{equation}
where
\begin{equation}
\label{eq:kappacool}
    \kappa^2 = k_z^2\left[ 1 - 4\left( \frac{\sigma \Omega + A\omega_\mathrm{Ai}/\!\sqrt{\mu \rho_\mathrm{i}}}{\sigma^2 - \omega_\mathrm{Ai}^2} \right)^2 \right]\left( 1 - \frac{\sigma^2}{\omega_\mathrm{Ai}^2} \right).
\end{equation}

As seen, Eq.~(\ref{eq:depm1}) is a Bessel equation for $p_{\mathrm{m}1} \equiv p_{\mathrm{t}1}$ and it solution inside the flux tube is
\begin{equation}
\label{eq:pt1inside}
    p_{\mathrm{t}1}(r) = \alpha_\mathrm{i}I_m(\kappa r) \quad \mbox{for} \quad r \leqslant a,
\end{equation}
where $I_m$ is the modified Bessel function of the first kind, and $\alpha_\mathrm{i}$ is a constant.

In similar way, starting from the basic equations of the incompressible ideal plasma \citep[e.g.,][]{Zaqarashvili2015}, one easily gets that the solutions for the total pressure perturbation in the environment is given by
\begin{equation}
\label{eq:pt1outside}
    p_{\mathrm{t}1}(r) = \alpha_\mathrm{e}K_m(k_z r) \quad \mbox{for} \quad r > a,
\end{equation}
where $K_m$ is the modified Bessel function of the second kind, and $\alpha_\mathrm{e}$ is a constant.

Following the standard method for deriving the wave dispersion relations, we have to merge the two solutions of $p_{\mathrm{t}1}(r)$ in both media at the tube boundary, $r = a$, via the boundary condition \citep{Zaqarashvili2015}
\begin{equation}
\label{eq:bc1}
    \left[ p_{\mathrm{t}1} + a\left( \rho_\mathrm{i}\Omega^2 - \frac{A^2}{\mu} \right)\xi_r \right]_a = 0,
\end{equation}
along with the condition for the continuity of the Lagrangian displacement $\xi_r$ across the boundary, that is,
\begin{equation}
\label{eq:bc2}
    \left[ \xi_r \right]_a = 0.
\end{equation}

The Lagrangian displacement, $\xi_r$, can be obtained from the $v_{1r}$ expression (\ref{eq:v1rfinal}) (and from a similar one for the case of incompressible plasma) through the simple relation
\[
    \xi_r = -\frac{v_{1r}}{\mathrm{i}\sigma}
\]
and one gets
\[
    \xi_r(r \leqslant a) = \frac{1}{\rho_\mathrm{i}}\frac{1}{Y}\frac{1}{\sigma^2 - \omega_\mathrm{Ai}^2}\left( \frac{\mathrm{d}}{\mathrm{d}r}p_{\mathrm{t}1} - Z\frac{m}{r}p_{\mathrm{t}1}  \right),
\]
\[
    \xi_r(r > a) = \frac{1}{\rho_\mathrm{e}}\frac{1}{\omega^2 - \omega_\mathrm{Ae}^2} \frac{\mathrm{d}}{\mathrm{d}r}p_{\mathrm{t}1},
\]
where the local Alfv\'en frequency in the environment is given by the expression
\[
    \omega_\mathrm{Ae} \equiv k_zB_\mathrm{e}/\!\sqrt{\mu \rho_\mathrm{e}} = k_z v_\mathrm{Ae}.
\]

Finally, by applying the boundary conditions (\ref{eq:bc1}) and (\ref{eq:bc2}), we arrive at the wave dispersion relation of the MHD modes propagating in the cool plasma--incompressible environment system, which in form coincides with Eq.~(42) in \citealp{Zaqarashvili2015}:
\begin{eqnarray}
\label{eq:dispeqc}
    \frac{\left( \sigma^2 - \omega_\mathrm{Ai}^2 \right)F_m(\kappa^\mathrm{c}_\mathrm{i}a) - 2m\left( \sigma \Omega + A\omega_\mathrm{Ai}/\! \sqrt{\mu \rho_\mathrm{i}} \right)}{\rho_\mathrm{i}\left( \sigma^2 - \omega_\mathrm{Ai}^2 \right)^2 - 4\rho_\mathrm{i}\left( \sigma \Omega + A\omega_\mathrm{Ai}/\! \sqrt{\mu \rho_\mathrm{i}} \right)^2} \nonumber \\
    \nonumber \\
    {}= \frac{P_m(k_z a)}{\rho_\mathrm{e}\left( \sigma^2 - \omega_\mathrm{Ae}^2 \right) - \left( \rho_\mathrm{i}\Omega^2 - A^2/\mu \right)P_m(k_z a)},
\end{eqnarray}
where
\[
    F_m(\kappa^\mathrm{c}_\mathrm{i}a) = \frac{\kappa^\mathrm{c}_\mathrm{i}aI_m^{\prime}(\kappa^\mathrm{c}_\mathrm{i}a)}{I_m(\kappa^\mathrm{c}_\mathrm{i}a)} \quad \mbox{and} \quad P_m(k_z a) = \frac{k_z aK_m^{\prime}(k_z a)}{K_m(k_z a)},
\]
but the wave attenuation coefficient in the internal medium, according to expression (\ref{eq:kappacool}), has the form
\[
    \kappa^\mathrm{c}_\mathrm{i} = k_z\left\{ 1 - 4\left( \frac{\sigma \Omega + A\omega_\mathrm{Ai}/\!\sqrt{\mu \rho_\mathrm{i}}}{\sigma^2 - \omega_\mathrm{Ai}^2} \right)^2 \right\}^{1/2}\left( 1 - \frac{\sigma^2}{\omega_\mathrm{Ai}^2}  \right)^{1/2}.
\]
Note, that the prime, $\prime$, in Eq.~(\ref{eq:dispeqc}) implies differentiation of the modified Bessel functions to their arguments.  The solutions to this dispersion relation are presented and discussed in the next section.

\section{Numerical solutions and results}
\label{sec:numerics}
For convenience in the numerical task, we normalize all velocities with respect to the Alfv\'en speed inside the flux tube, $v_\mathrm{Ai}$, and all lengths with respect to the tube radius, $a$.  As usual, we shall look for solutions of the wave phase velocity $v_\mathrm{ph} = \omega/k_z$ as a function of the axial wavenumber $k_z$, which in dimensionless variables reads as $v_\mathrm{ph}/v_\mathrm{Ai} = f(k_z a)$.  Since we expect, at some conditions, the occurrence of instability in the studied jet--environment system, it is logical to assume that the angular wave frequency $\omega$ is a complex quantity while the axial wavenumber $k_z$ is a real quantity.  This implies that the normalized wave phase velocity becomes complex number whose real part yields the wave phase velocity and its imaginary part gives the instability growth rate, both as functions of the dimensionless wavenumber $k_z a$.  The normalization of Alfv\'en local frequencies, the Doppler-shifted frequency, as well as the Alfv\'en speed in the surrounding coronal plasma requires the usage of the two twist parameters $\varepsilon_1$, $\varepsilon_2$, and the magnetic fields ratio $b = B_\mathrm{e}/B_{\mathrm{i}z}$, respectively.  In addition, the dimensionless axial flow velocity is presented by the Alfv\'en Mach number $M_\mathrm{A} = U_z/v_\mathrm{Ai}$.  To sum up, the input parameters at each run of the code solving the transcendental dispersion equation in complex variables are: $m$, $\eta$, $\varepsilon_1$, $\varepsilon_2$, $b$, and $M_\mathrm{A}$.  In \citet{Zaqarashvili2015} it has been shown that the instability in an untwisted rotating flux tube at sub-Alfv\'enic jet velocities can occur if
\begin{equation}
\label{eq:criterion}
    \frac{a^2 \Omega^2}{v_\mathrm{Ai}^2} > \frac{1 + \eta}{1 + |m|\eta}\,\frac{(k_z a)^2}{|m| - 1}(1 + b^2).
\end{equation}
This inequality is also applicable to twisted magnetic flux tubes at small values of $\varepsilon_1$, say between $0.001$ and $0.005$.  Let us assume that the axial speed at which the jet propagates is the critical speed for arising the KHI.  In such a way we define the magnitude of the Alfv\'en Mach number -- its values for the two moving tubes with different magnetic field twist parameters will be specified later on.  The above inequality can be rearranged in the form
\begin{equation}
\label{eq:instcond}
    (k_z a)_\mathrm{rhs} < \left[ \left( \frac{U_\phi}{v_\mathrm{Ai}} \right)^2 \frac{1 + |m|\eta}{1 + \eta}\,\frac{|m| - 1}{1 + b^2} \right]^{1/2},
\end{equation}
which defines the right-hand limit of the instability region, that is, instability can arise for all $k_z a$ less than some number (the right-hand side of inequality (\ref{eq:instcond})) (recall that $U_\phi = \Omega a$ is the rotating velocity of the jet).  On the other hand, one can talk for instability if the unstable wavelength is shorter than the height of the jet and this requirement defines the left-hand limit of the instability region:
\begin{equation}
\label{eq:lhlimit}
    (k_z a)_\mathrm{lhs} > \frac{\pi \Delta \ell}{H}.
\end{equation}
For our jet this limit is equal to $0.524$.  The right-hand limit, as seen from inequality (\ref{eq:instcond}), depends on the density contrast, $\eta$, magnetic fields ratio, $b$, and on azimuthal mode number, $m$.  One can expect that inequality (\ref{eq:instcond}) will give at $\varepsilon_1 = 0.005$ an indicative value of $(k_z a)_\mathrm{rhs}$.  Such an estimation of the width of the instability range/window was used by us \citep{Zhelyazkov2018} in studying the KHI in the coronal hole jet observed by \citet{Chen2012}.  Computations showed that at small MHD mode numbers (equal to $2$ and $3$, respectively) the shortest wavelength at which the instability manifests itself is of $85$~Mm -- a value which was not comfortable for observed KH blobs.  From a physical point of view, it is more logical to expect that wavelength to be of the order of the interspace between the moving blobs, or roughly speaking of the order of the half width of the jet.  A reliable azimuthal mode number at which that happens for the \citet{Chen2012} jet, as we already discussed in the Introduction section, is $m = 12$.

Some preliminary computations for our jet via its two presentations with $\varepsilon_1 = 0.005$ and $0.1$ show that one can observe unstable waves with wavelengths equal or longer than $65$~Mm at $m = 3$ for the magnetic field twist of $0.005$ and at $m = 4$ for $\varepsilon_1 = 0.1$.  In other words, the intermediately twisted flux tube requires a higher mode number to have approximately the same width of the instability window as that of the slightly twisted one.  The shortest wavelength that we chose for comparison was $\lambda_\mathrm{KH} = 65$~Mm.  The instability developing/growth times at that wavelength in the two flux tubes were ${\cong}16.4$~min for $\varepsilon_1 = 0.005$ and ${\cong}7.1$~min at $\varepsilon_1 = 0.1$.  The two growth times are shorter than the jet lifetime of $30$~min and the KHI can, in principle, develop in both flux tubes, respectively.  This wavelength, however, is still too long to be comparable with the jet's width $\Delta \ell \cong 30$~Mm.  Obviously we should look for higher MHD modes with wider instability ranges, within which the KHI with wavelengths of, say, $12$ and $15$~Mm would occur.  The dimensionless wavenumbers that correspond to these wavelengths are $7.854$ and $6.283$, respectively.  Thus, we have to find those $m_{\epsilon_1 = 0.005}$ and $m_{\epsilon_1 = 0.1}$, which will ensure instability windows with widths of the order of $8.5$--$9$.  A rough estimation of the azimuthal mode number for the $\varepsilon_1 = 0.005$-flux tube can be obtained by rearranging the instability criterion (\ref{eq:criterion}) in the form
\begin{equation}
\label{eq:findingm}
    \eta |m|^2 + (1 - \eta)|m| - 1 - \frac{(k_z a)^2(1 + \eta)(1 + b^2)}{(U_\phi/v_\mathrm{Ai})^2} > 0.
\end{equation}
With $\eta = 0.864$, $k_z a = 7.854$, $U_\phi = 180$~km\,s$^{-1}$, and $b_{\epsilon_1 = 0.005} = 0.679$, the above equation yields $m = 18$.  This magnitude is, however, overestimated -- the numerical calculations show that the appropriate MHD wave mode number that accommodates the unstable wavelength of $12$~Mm ($k_z a = 7.854$) is $m_{\epsilon_1 = 0.005} = 16$.  The higher mode number for the $\varepsilon_1 = 0.1$-flux tube which will give approximately the same instability window should be guessed.  The computations yielded that $m_{\epsilon_1 = 0.1} = 18$ fits the bill -- this mode number ironically coincides with that $m$ which was estimated from inequality (\ref{eq:findingm}) applied to the flux tube with magnetic field twist parameter $\varepsilon_1 = 0.005$.  In the next two subsections we present and comment on the results of the numerics for these two mode numbers.

\subsection{Instability characteristics of the $m = 16$ MHD mode in a flux tube with $\epsilon_1 = 0.005$}
\label{subsec:eta0464}
\begin{figure}
\centering
  \begin{minipage}{\columnwidth}
  \centering
    \includegraphics[width=7.5cm]{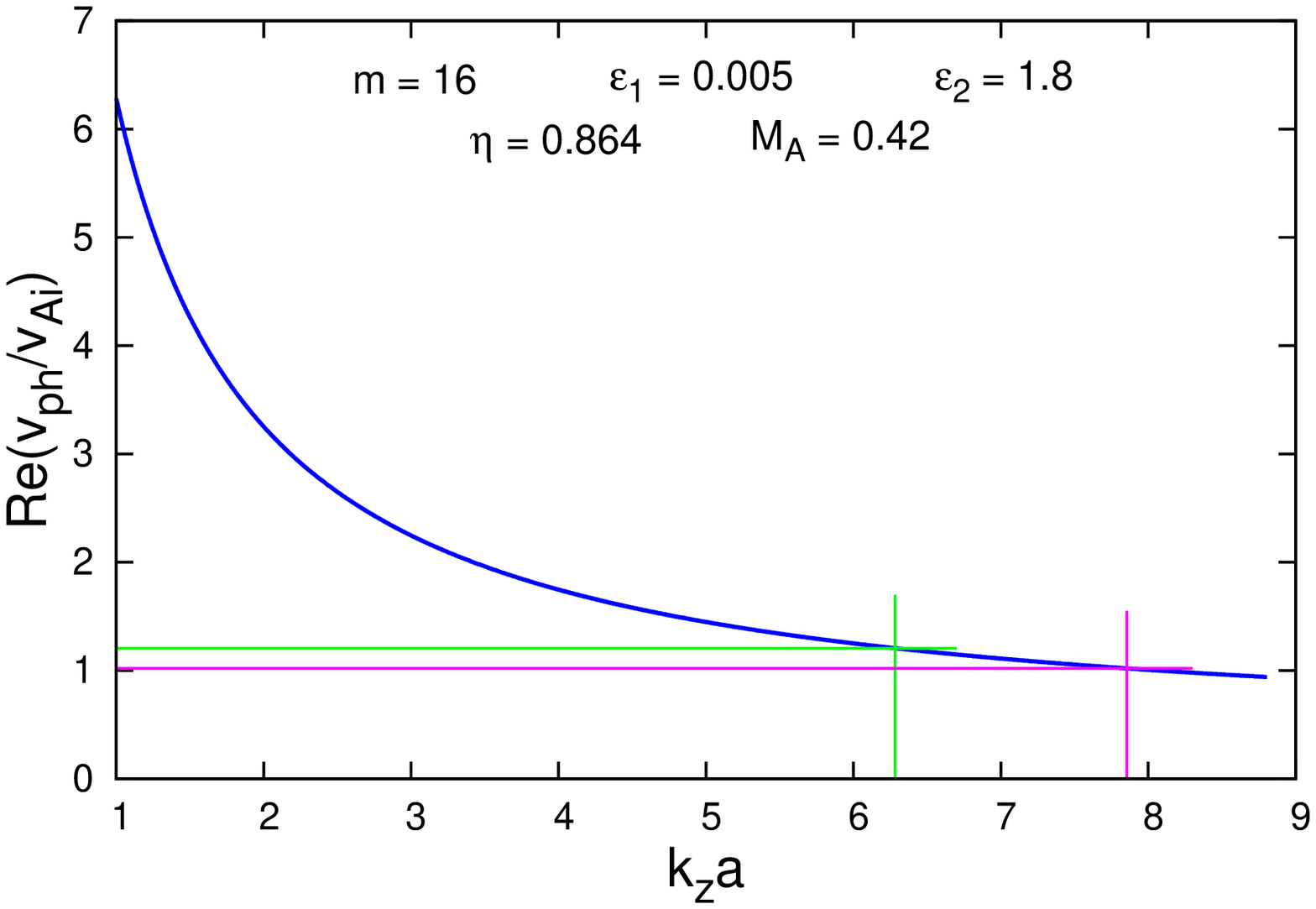} \\
\vspace{0mm}
    \includegraphics[width=7.5cm]{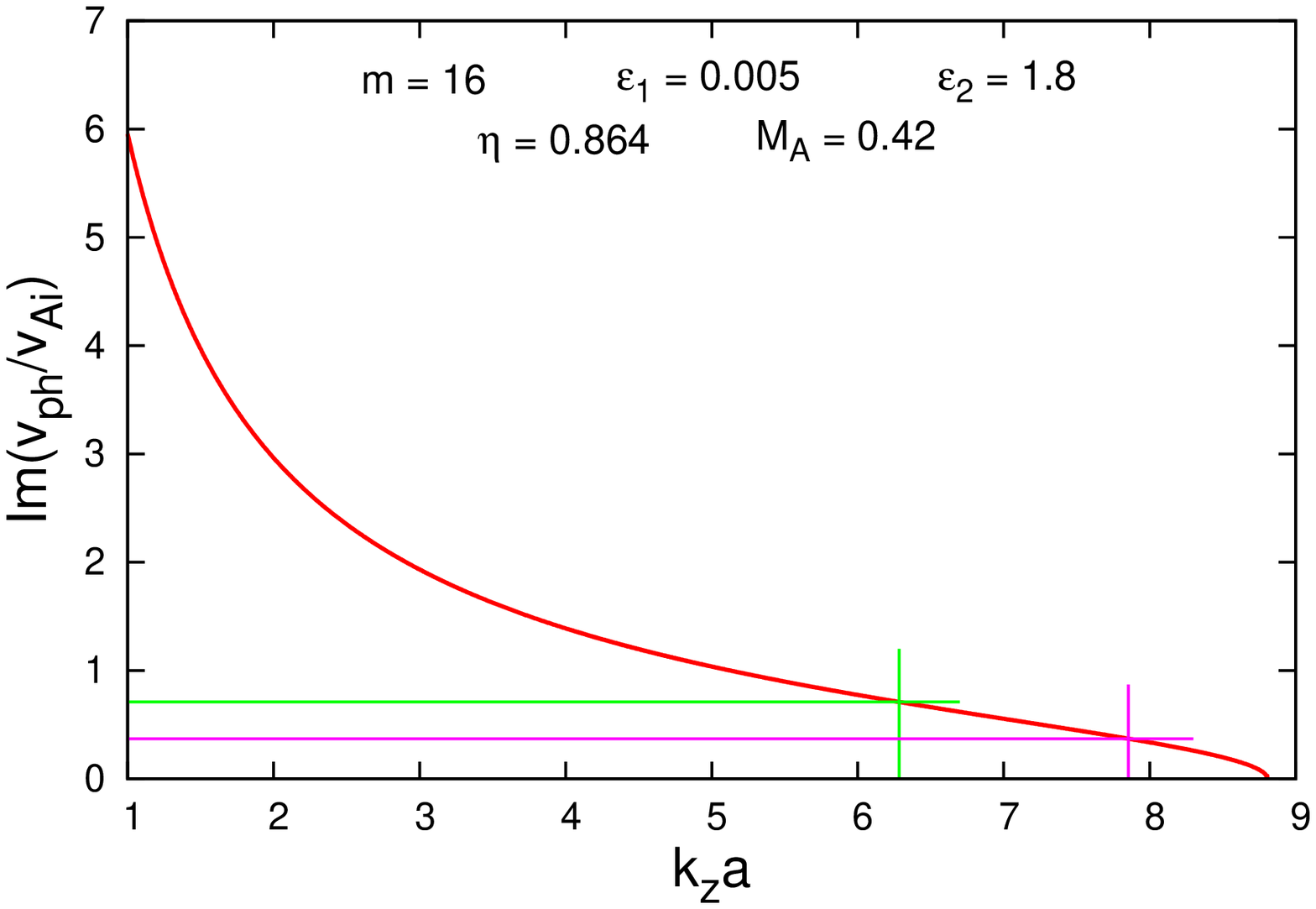}
  \end{minipage}
  \caption{Dispersion curve of the unstable $m = 16$ MHD mode propagating along a twisted cool magnetic flux tube at $\eta = 0.864$, $b = 0.679$, $M_{\rm A} = 0.42$, $\varepsilon_1 = 0.005$, and $\varepsilon_2 = 1.8$.  (\emph{Bottom panel}) Normalized growth rate curve of the unstable $m = 16$ MHD mode propagating along a twisted cool magnetic flux tube at the same parameters as in the top panel.}
   \label{fig:fig2}
\end{figure}

The input parameters in the numerical task of solving Eq.~(\ref{eq:dispeqc}) are: $m = 16$, $\eta = 0.864$, $b = 0.679$, $\varepsilon_1 = 0.005$, $\varepsilon_2 = 1.8$, and $M_\mathrm{A} = 0.42$.  The results are pictured in Fig.~\ref{fig:fig2}.
The KHI characteristics, namely the wave growth rate, $\gamma_\mathrm{KH}$, growth time, $\tau_\mathrm{KH} = 2\pi/\gamma_\mathrm{KH}$, and wave velocity, $v_\mathrm{ph}$, calculated from the graphics in Fig.~\ref{fig:fig2}, for the two wavelengths are as follows:

At $\lambda_\mathrm{KH} = 12$~Mm we have
\[
    \gamma_\text{KH} \cong 45.65 \times 10^{-3}\:\text{s}^{-1}, \;\, \tau_\text{KH} \cong 2.3\:\text{min}, \;\, v_\text{ph} \cong 240\:\text{km}\,\text{s}^{-1},
\]
while at $\lambda_\mathrm{KH} = 15$~Mm we have
\[
    \gamma_\text{KH} \cong 70.0 \times 10^{-3}\:\text{s}^{-1}, \;\, \tau_\text{KH} \cong 1.5\:\text{min}, \;\, v_\text{ph} \cong 284\:\text{km}\,\text{s}^{-1}.
\]
As is seen, the wave phase velocities are supper-Alfv\'enic.  One specific property of instability $k_z a$ ranges of a rotating magnetic flux tube is that for a fixed MHD mode number $m$ its width depends upon the magnetic field twist parameter $\varepsilon_1$.  With increasing the value of $\varepsilon_1$, the instability window becomes narrower and at some critical magnetic field twist its width equals zero.  This circumstance implies that for $\varepsilon_1 \geqslant \varepsilon_1^\mathrm{cr}$ there is no instability, or, in other words, there exists a critical azimuthal magnetic field $B_{\phi}^\mathrm{cr} = \varepsilon_1^\mathrm{cr} B_{\mathrm{i}z}$ that suppresses the instability onset.  In the next Fig.~\ref{fig:fig3}, a series of dispersion and dimensionless
\begin{figure}
\centering
  \begin{minipage}{\columnwidth}
  \centering
    \includegraphics[width=7.5cm]{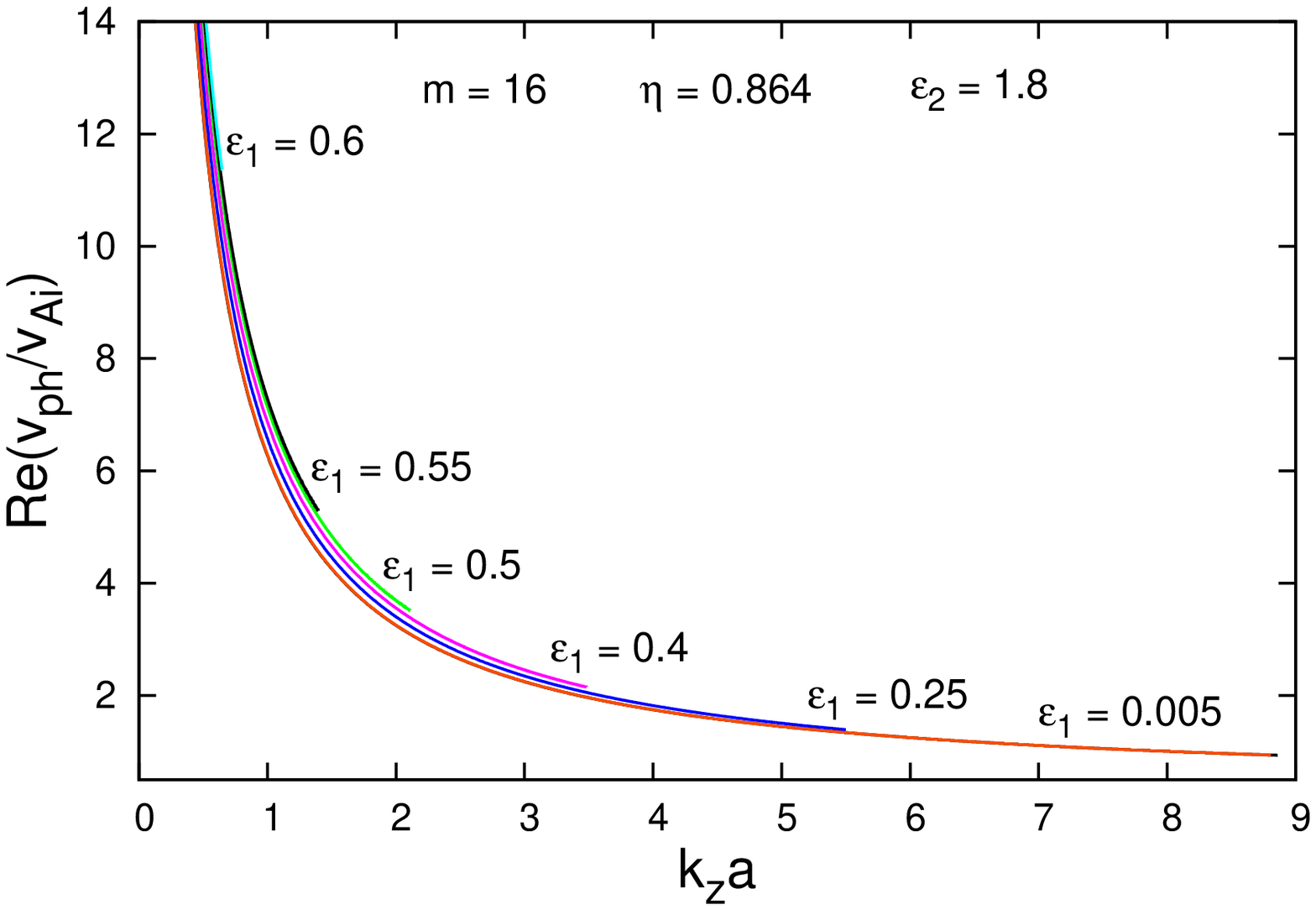} \\
\vspace{0mm}
    \includegraphics[width=7.5cm]{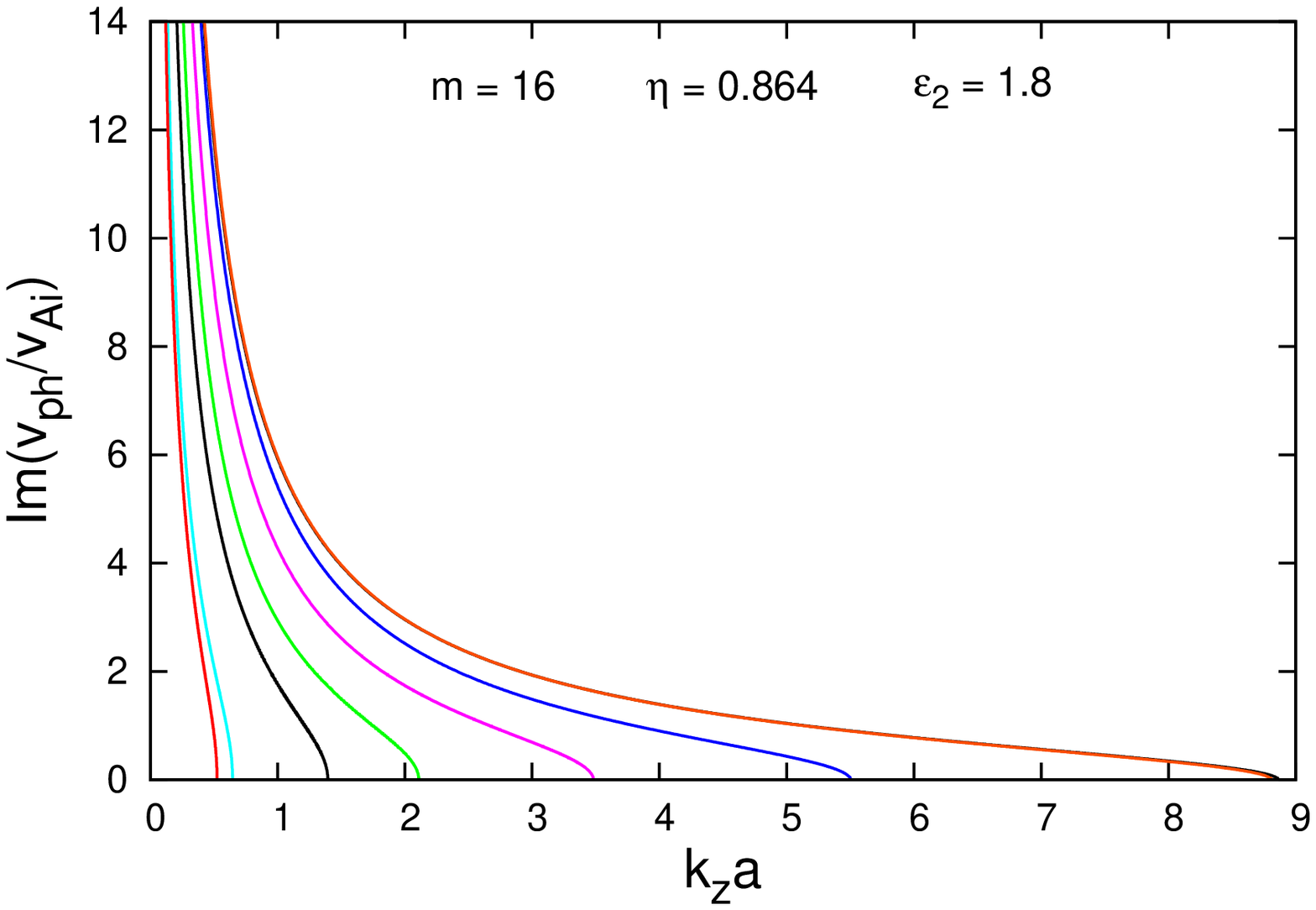}
  \end{minipage}
  \caption{(\emph{Top panel})Dispersion curves of the unstable $m = 16$ MHD mode propagating along a twisted cool magnetic flux tube at $\eta = 0.864$, $b = 0.679$, $\varepsilon_2 = 1.8$, and the following values of $\varepsilon_1$ (from right to left): $0.001$ (black curve, coinciding with the orange one), $0.005$, $0.25$, $0.4$, $0.5$, $0.55$, $0.6$, and $0.60763$.  Alfv\'en Mach numbers for these curves are respectively $0.42$, $0.42$, $0.44$, $0.46$, $0.47$, $0.48$, $0.49$ and $0.4968$.  (\emph{Bottom panel}) Growth rates of the unstable $m = 16$ mode for the same input parameters.  The azimuthal magnetic fields that corresponds to the red $\varepsilon_1 = 0.60763$ curve (the instability window with zero width) is equal to $3.8$~G.  Real/observable $m = 16$ unstable MHD modes can be detected for $\varepsilon_1 < 0.60763$, which means that an azimuthal magnetic field of $3.8$~G would suppress the KHI onset.}
   \label{fig:fig3}
\end{figure}
wave phase velocity growth rates for various increasing magnetic field twist parameter's values has been plotted.  Note that each larger $\varepsilon_1$ implies an increase in $B_{\phi}$ and respectively a decrease of $B_{\mathrm{i}z}$ which automatically requires a multiplication of the initial $b$ and $M_\mathrm{A}$ values by $\sqrt{ 1 + \varepsilon_1^2}$ \citep{Zhelyazkov2016}.  The red dispersion curve in the bottom panel of Fig.~\ref{fig:fig3} has been obtained for $\varepsilon_1^\mathrm{cr} = 0.60763$ and visually defines the left-hand limit of all the other instability ranges.  The azimuthal magnetic field that stops the KHI is equal to $3.8$~G.

\subsection{Instability characteristics of the $m = 18$ MHD mode in a flux tube with $\epsilon_1 = 0.1$}
\label{subsec:eta0864}
\begin{figure}
\centering
  \begin{minipage}{\columnwidth}
  \centering
    \includegraphics[width=7.5cm]{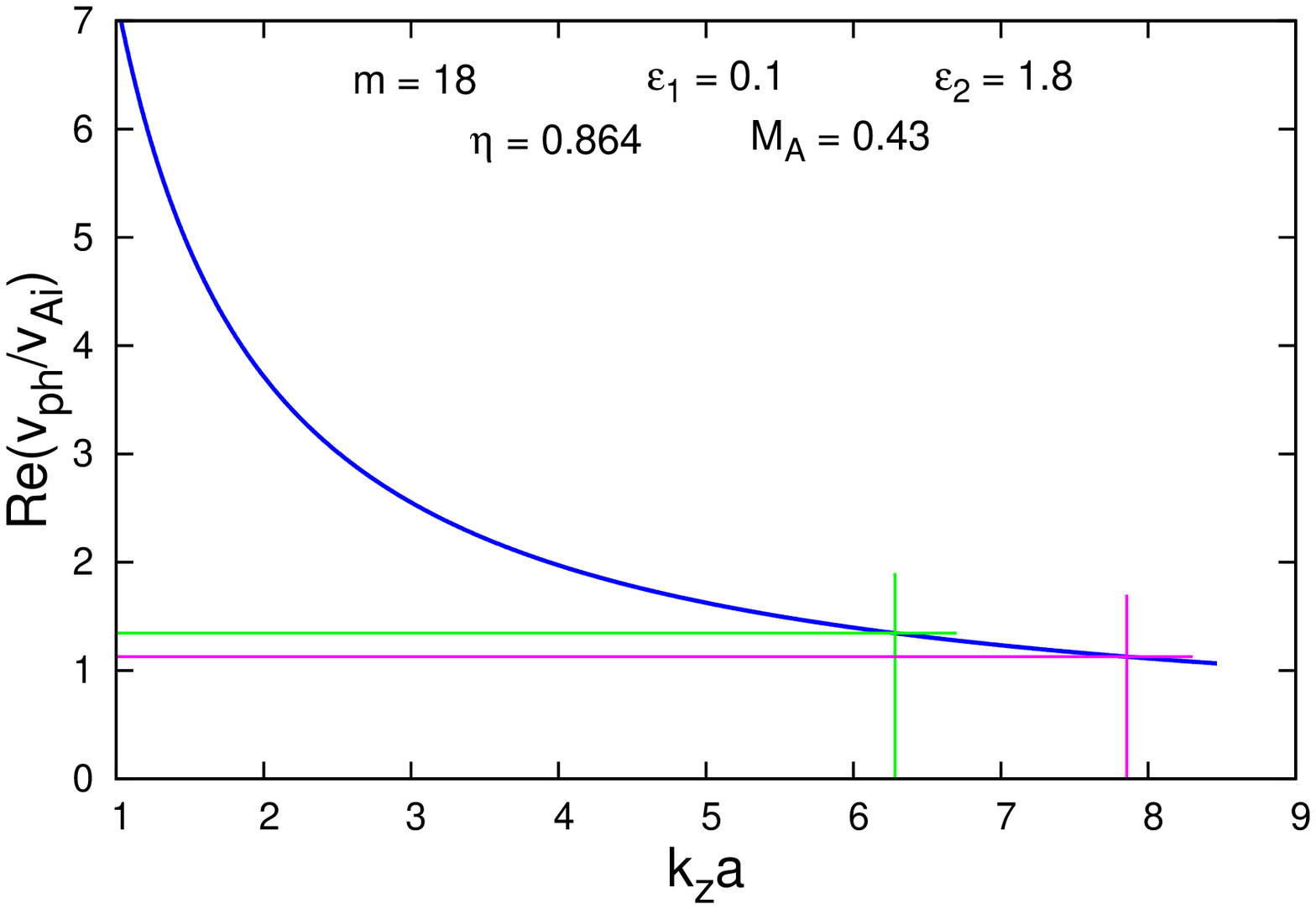} \\
\vspace{0mm}
    \includegraphics[width=7.5cm]{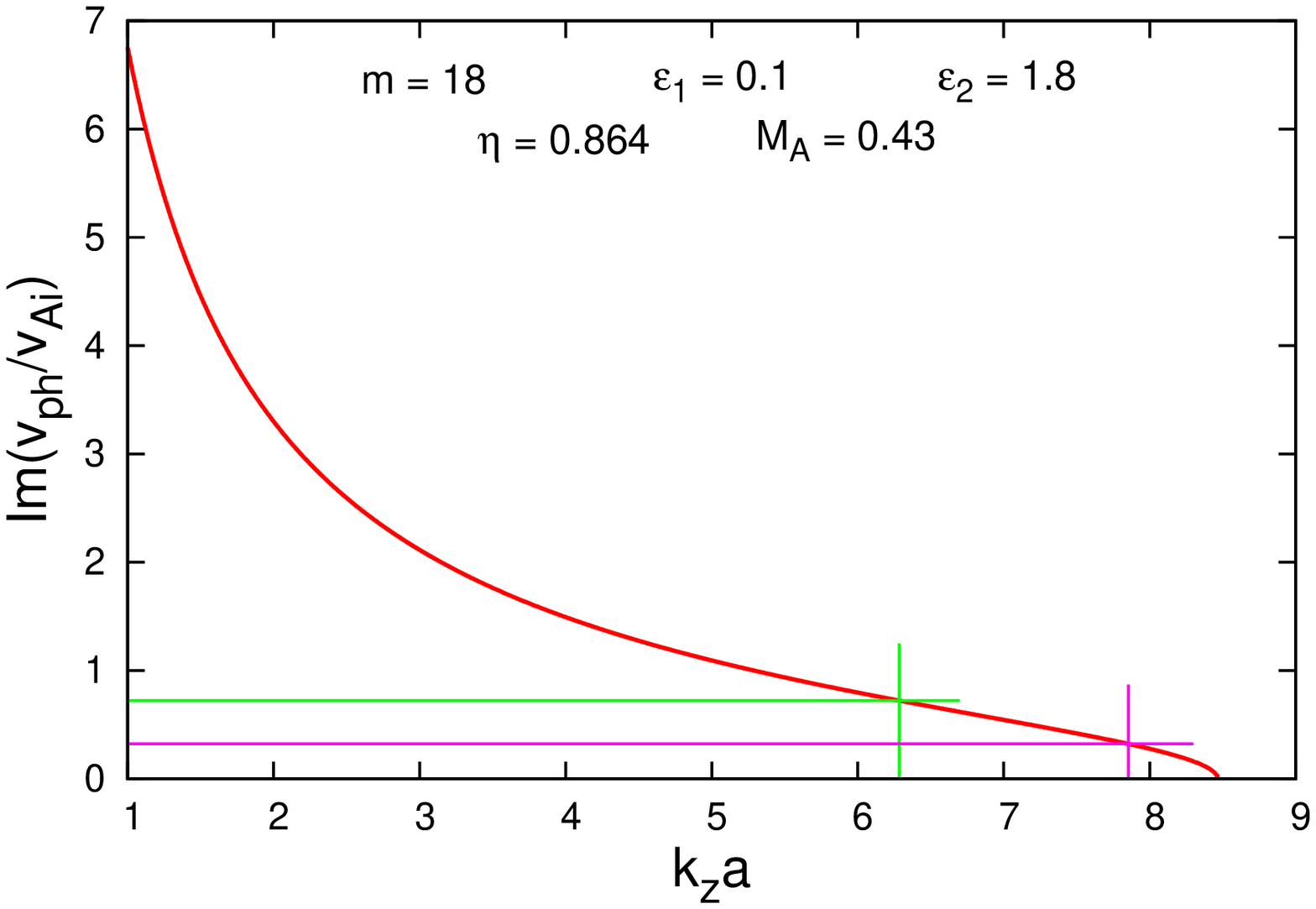}
  \end{minipage}
  \caption{(\emph{Top panel}) Dispersion curve of the unstable $m = 18$ MHD mode propagating along a twisted cool magnetic flux tube at $\eta = 0.864$, $b = 0.682$, $M_{\rm A} = 0.43$, $\varepsilon_1 = 0.1$, and $\varepsilon_2 = 1.8$.  (\emph{Bottom panel}) Normalized growth rate curve of the unstable $m = 18$ MHD mode propagating along a twisted cool magnetic flux tube at the same parameters as in the top panel.}
   \label{fig:fig4}
\end{figure}

\begin{figure}
\centering
  \begin{minipage}{\columnwidth}
  \centering
    \includegraphics[width=7.5cm]{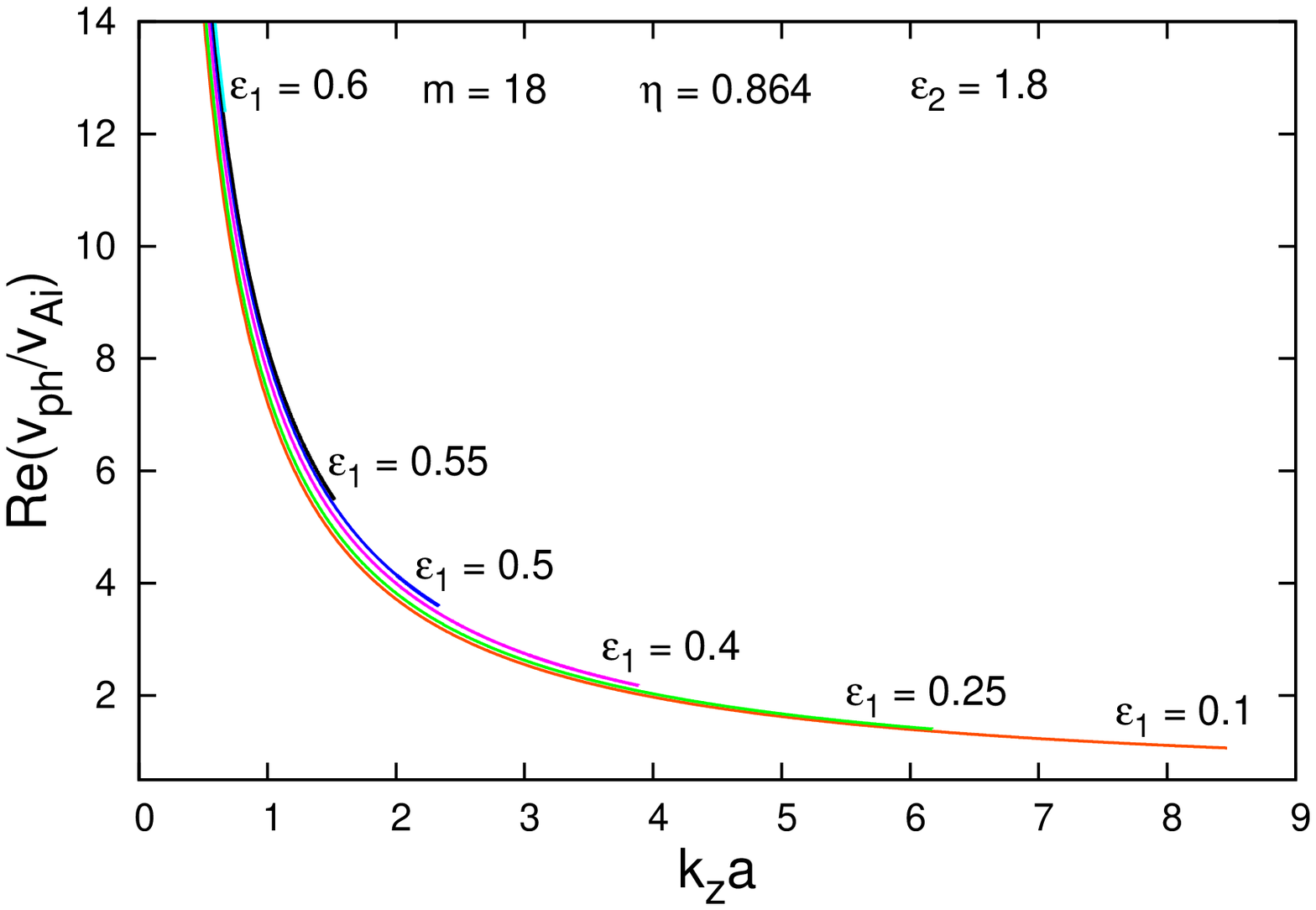} \\
\vspace{0mm}
    \includegraphics[width=7.5cm]{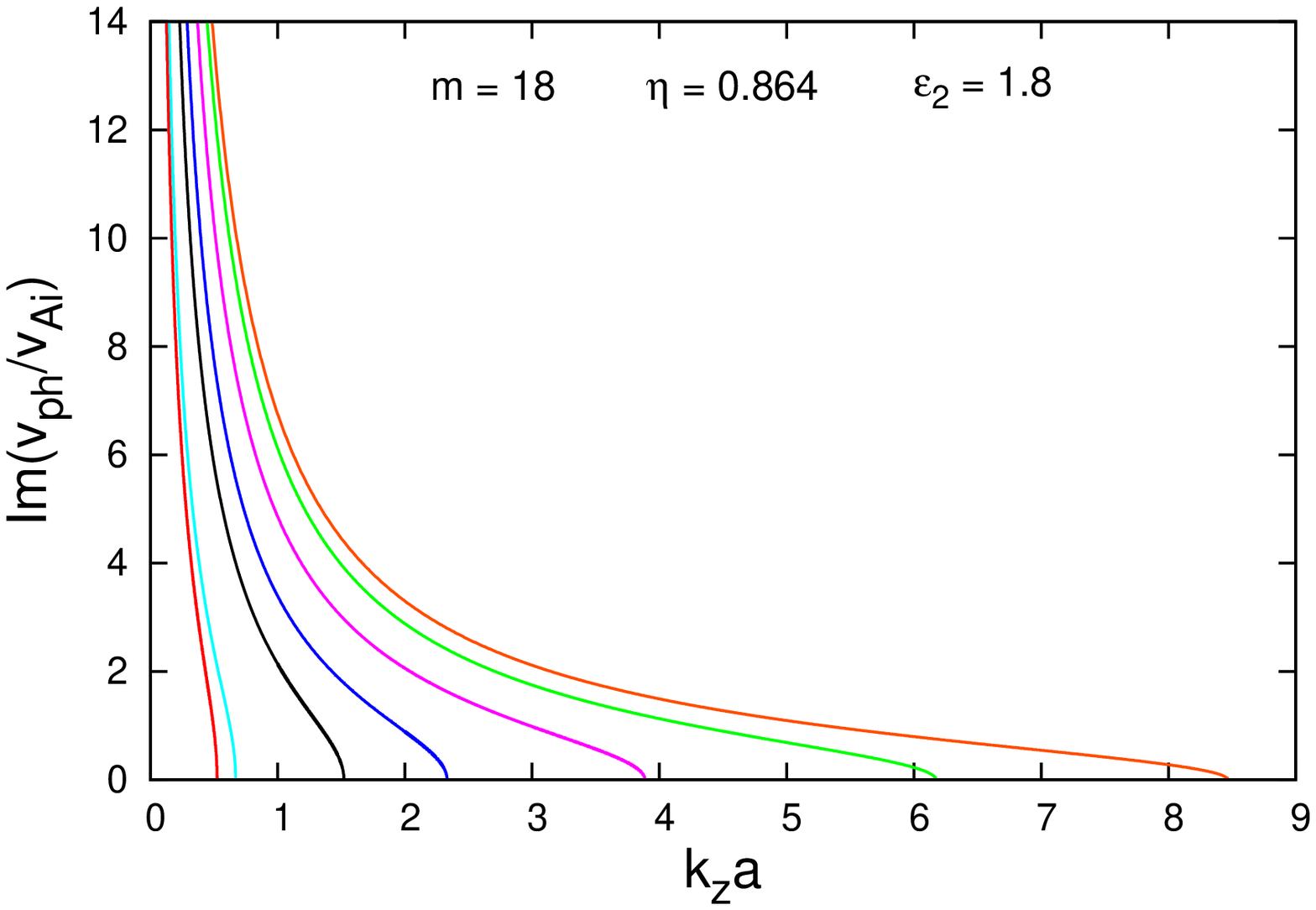}
  \end{minipage}
  \caption{(\emph{Top panel})Dispersion curves of the unstable $m = 18$ MHD mode propagating along a twisted cool magnetic flux tube at $\eta = 0.864$, $b = 0.682$, $\varepsilon_2 = 1.8$, and the following values of $\varepsilon_1$ (from right to left): $0.1$, $0.25$, $0.4$, $0.5$, $0.55$, $0.6$, and $0.608049$.  Alfv\'en Mach numbers for these curves are respectively $0.43$, $0.44$, $0.46$, $0.47$, $0.48$, $0.49$ and $0.4969$.  (\emph{Bottom panel}) Growth rates of the unstable $m = 18$ mode for the same input parameters.  The azimuthal magnetic fields that corresponds to the red $\varepsilon_1 = 0.608049$ curve (the instability window with zero width) is equal to $3.8$~G.  Real/observable $m = 18$ unstable MHD modes can be detected for $\varepsilon_1 < 0.608049$, which means that an azimuthal magnetic field of $3.8$~G would suppress the KHI onset.}
   \label{fig:fig5}
\end{figure}

With $m = 18$, $\eta = 0.864$, $b = 0.682$, $\epsilon_1 = 0.1$, $\epsilon_2 = 1.8$, and $M_\mathrm{A} = 0.43$ the numerical code yields curves for the unstable $m = 18$ MHD mode that are plotted in Fig.~\ref{fig:fig4}.  It is very surprising that we have obtained similar growth times as those of the
$m = 16$ MHD mode.  The values extracted from the plots are as follows:

At $\lambda_\mathrm{KH} = 12$~Mm one obtains
\[
    \gamma_\text{KH} \cong 39.48 \times 10^{-3}\:\text{s}^{-1}, \;\, \tau_\text{KH} \cong 2.6\:\text{min}, \;\, v_\text{ph} \cong 264\:\text{km}\,\text{s}^{-1},
\]
while at $\lambda_\mathrm{KH} = 15$~Mm we have
\[
    \gamma_\text{KH} \cong 70.89 \times 10^{-3}\:\text{s}^{-1}, \;\, \tau_\text{KH} \cong 1.5\:\text{min}, \;\, v_\text{ph} \cong 315\:\text{km}\,\text{s}^{-1}.
\]
As before, the phase velocities of the unstable $m = 18$ MHD mode are super-Alfv\'enic.  The dependence of the instability range on the value of the magnetic field twist parameter also is similar, as it can be seen from Fig.~\ref{fig:fig5}.
The most striking result is the circumstance that both MHD modes yield the same suppressing the KHI onset azimuthal magnetic field of $3.8$~G.  This is due to the fact that both critical magnetic field twist parameters, $\varepsilon_1^{\mathrm{cr}}$s, are of the same order.

\section{Summary and conclusion}
\label{sec:summary}
In this article, we have studied how the condition for occurrence of unstable MHD modes propagating on a twisted rotating jet emerging from a filament eruption depends upon the degree of the jet's magnetic field twist.  We model the jet as a moving rotating twisted magnetic flux tube of radius $a$ with homogeneous density $\rho_\mathrm{i}$ through a coronal plasma with homogeneous density $\rho_\mathrm{e}$ embedded in a homogeneous magnetic field.  The magnetic field twist is characterized by the numerical parameter $\epsilon_1 \equiv B_\phi/B_z$, where the azimuthal and the axial magnetic field components are evaluated at the tube boundary, $r = a$.  We consider two cases associated with the internal magnetic field twist, namely a slightly twisted flux tube, with $\epsilon_1 = 0.005$, and an intermediately twisted one, for which $\epsilon_1 = 0.1$.  The density contrast between the jet and its environment is defined as $n_\mathrm{e}/n_\mathrm{i}$.  The pressure balance equation that relates the internal and external sound and Alfv\'en speeds at fixed rotating speed, plasma density contrast, and environment's magnetic field, yields different jet's Alfv\'en speeds for the aforementioned two magnetic field twist parameters.  Measured physical parameters of the jet and its environment (plasma densities, electron temperatures, rotating speed, and so on), at an assumed external magnetic field of $5$~G yield a jet's plasma beta less than $1$ (${=}0.596$), while that of the environment is bigger than $1$ (${=}1.196$).  This means that we can treat the jet as a cool medium while its environment can be considered as quasi incompressible plasma \citep{Zank1993}.  Under these circumstances, for modeling the wave propagation in the jet--environment system, it was necessary to derive a new dispersion equation governing the propagation of the excited MHD modes, the latter being a complement to the dispersion equation of high MHD modes obtained in \citealp{Zaqarashvili2015} under the assumption that the two media are incompressible plasmas.

The numerical solutions (in complex variable when looking for unstable waves) to the dispersion equation show that there exist finite-sized $k_z a$ regions, in which the propagating MHD mode becomes unstable and the instability that emerges is of the Kelvin--Helmholtz kind.  The width of the instability region/window depends on a few parameters, notably the density contrast, the magnetic fields ratio (external magnetic field over the internal one), the ratio of rotating jet speed and internal Alfv\'en speed, the MHD mode number $m$, and the twist of the magnetic field $\varepsilon_1$ (we must emphasize that KHI can exist in an untwisted rotating magnetic flux tube -- the instability window at $\varepsilon_1 = 0$ practically coincides with that for $\varepsilon_1 = 0.001$).  At fixed density contrast and jet's rotating speed, the rest two parameters (magnetic fields ratio, $b \equiv B_\mathrm{e}/B_{\mathrm{i}z}$, and Alfv\'en Mach number, $M_\mathrm{A} \equiv U_z/v_\mathrm{Ai}$) are different for the two choices of the magnetic field twist.  To make a comparison between the two presentation of the jet, we have decided to see what instability characteristics would be obtained at two fixed wavelengths equal to $12$ and $15$~Mm, respectively (both being comparable to the half width of the jet (see Fig.~\ref{fig:fig1}).  It has been found that two approximately equal in width instability regions, which should accommodate the two wavelengths (that is, their respective $k_z a$ values), can be achieved if in the case of slightly twisted flux tube ($\varepsilon_1 = 0.005$) the MHD mode number is $m = 16$ while at the moderately twisted tube ($\varepsilon_1 = 0.1$) the mode number must be $m = 18$.

It is rather surprising that the modes instability growth times (defined as $2\pi$ divided by their instability growth rates) turned out to be of the same order, $2.3\;\&\;2.6$~min, at $\lambda_\mathrm{KH} = 12$~Mm, and $1.5$~min at $\lambda_\mathrm{KH} = 15$~Mm, respectively.  Both instability growth times seems reasonable bearing in mind that the jet's life time was ${\approx}30$~min.  In both jet's presentations, the phase velocities of unstable $m = 16$ and $m = 18$ MHD modes are slightly supper-Alfv\'enic.

The width of the instability $k_z a$ range depends very sensitively on magnetic field twist parameter.  If for its small values in the range of $0.001$--$0.005$ the instability windows are almost the same in size (see Fig.~\ref{fig:fig2}), with the increase in the magnetic twist the instability range quickly becomes narrower and at some critical magnetic field twist $\varepsilon_1^\mathrm{cr}$ its width equals zero, that is, the instability is stopped.  The critical magnetic field twist weakly depends on the MHD mode number: for $m = 16$ its magnitude is $0.60763$, while at $m = 18$ it is equal to $0.608049$.  These numbers help us to find out those azimuthal magnetic field components that suppress KHI onset \citep{Zaqarashvili2015} -- their magnitudes for both modes, as can anticipate bearing in mind the very close values of the critical magnetic field twist parameters, are the same, that is, equal to $3.8$~G.

As it is seen from Table~\ref{tab:parameter}, the electron number densities and temperatures are displayed with their estimation errors.  An immediate question that raises is whether any  small changes in $n$ and $T$ in the limits of their estimation errors will change significantly the conditions for the KHI occurrence.  Generally speaking, the strongest influence on the instability parameters are the variations of the density contrast and partially those of the electron temperatures.  An additional investigation of that effect in an intermediately twisted moving flux tube with $\varepsilon_1 = 0.1$ shows that even at the biggest and smallest density contrasts that can be defined from the Table~\ref{tab:parameter}'s values, correspondingly equal to $0.845$ and $0.884$, the frequency growth rates, growth times, and phase velocity of the unstable $m = 18$ mode at two wavelengths found at $\eta = 0.845$ and $0.884$ are very close to the values obtained at $\eta = 0.864$ (see Table~\ref{tab:khidata}).  A slight increase/decrease by $0.01$--$0.02$~K in the two electron temperatures do not change noticeably plasma betas and consequently the instability characteristics shown in Table~\ref{tab:khidata} remain practically unchanged.

\begin{table}
\caption{Kelvin--Helmholtz instability characteristics of the $\varepsilon_1 = 0.1$-flux tube at three different density contrasts.}
\label{tab:khidata}
\vspace*{2mm}

\centering

\begin{tabular}{cccc}
\hline
$\eta$ & $\gamma_\mathrm{KH}$ & $\tau_\mathrm{KH}$ & $v_\mathrm{ph}$ \\
       & (${\times}10^{-3}$ s$^{-1}$)  &  (s) & (km\,s$^{-1}$) \\
\hline
\multicolumn{4}{c}{$\lambda_\mathrm{KH} = 12$ Mm} \\ \hline
$0.845$ & $38.61$ & $163$ & $267$  \\
$0.864$ & $39.48$ & $159$ & $264$  \\
$0.884$ & $38.33$ & $164$ & $260$  \\
\hline
\multicolumn{4}{c}{$\lambda_\mathrm{KH} = 15$ Mm} \\ \hline
$0.845$ & $70.58$ & $89.0$ & $318$  \\
$0.864$ & $70.89$ & $88.6$ & $264$  \\
$0.884$ & $70.46$ & $89.2$ & $312$  \\
\end{tabular}
\end{table}

Although the two jet's presentations yield similar growth times and wave phase velocities, we do think that the rotating and axially moving flux tube with a magnetic field twist of $0.1$ should adequately model the KHI in the jet observed on 2013 April 10--11.  Our way of modeling KHI in rotating solar jets is rather flexible and it can be applied to investigating any tornado-like event in the solar atmosphere provided that some basic jet's and instability parameters are available from the observations.

Our modeling of Kelvin--Helmholtz instability of high MHD modes in a twisted rotating and axially moving solar jet naturally needs some improvement, mostly in taking into account the radial jet's plasma density inhomogeneity that introduces such effects as the continuous spectra and resonant wave absorption \citep{Goedbloed2004}, which can modify in some extent the results obtained here.

\section*{Acknowledgements}

The work of I.Zh.\ and R.C.\ was supported by the Bulgarian Science Fund and the Department of Science \& Technology, Government of India Fund under Indo--Bulgarian bilateral project DNTS/INDIA 01/7, /Int/Bulgaria/P-2/12.  The authors are thankful to the \emph{Solar Dynamics Observatory}, the data from which are used in the present investigation, as well as to the reviewer for constructive criticisms and valuable comments, which were of great help in revising the manuscript.





\begin{thebibliography}{99}

\bibitem[\protect\citeauthoryear{Aschwanden et al.}{2013}]{Aschwanden2013}
    Aschwanden, M.~J., Boerner, P., Schrijver, C.~J., Malanushenko, A., 2013, Sol.\ Phys., 238, 5

\bibitem[\protect\citeauthoryear{Bennett \& Erd\'elyi}{2015}]{Bennett2015}
    Bennett, S.~M., Erd\'elyi, R., 2015, ApJ, 808, 135

\bibitem[\protect\citeauthoryear{Chandrasekhar}{1961}]{Chandrasekhar1961}
    Chandrasekhar, S., 1961, \emph{Hydrodynamic and Hydromagnetic Stability}, Clarendon Press, Oxford, Chap.~11

\bibitem[\protect\citeauthoryear{Chen et al.}{2012}]{Chen2012}
    Chen, H.-D., Zhang, J., Ma, S.-L., 2012, Res.\ Astron.\ Astrophys., 12, 573

\bibitem[\protect\citeauthoryear{Cranmer et al.}{2015}]{Cranmer2015}
    Cranmer, S.~R., Asgari-Targhi, M., Miralles, M.~P., Raymond, J.~C., Strachan, L., Tian, H., Woolse, L.~N., 2015, Philos.\ Trans.\ R.\ Soc.\ A, 373, 20140148

\bibitem[\protect\citeauthoryear{Curdt \& Tian}{2011}]{Curdt2011}
    Curdt, W., Tian, H., 2011, A\&A, 532, L9

\bibitem[\protect\citeauthoryear{De Pontieu et al.}{2012}]{DePontieu2012}
    De Pontieu, B., Carlsson, M., Rouppe van der Voort, L.~H.~M., Rutten, R.~J., Hansteen, V.~H., Watanabe, H., 2012, ApJ, 752, L12

\bibitem[\protect\citeauthoryear{Filippov et al.}{2015}]{Filippov2015}
    Filippov, B., Srivastava, A.~K., Dwivedi, B.~N., Masson, S., Aulanier, G., Joshi, N.~C., Uddin, W., 2015, MNRAS, 451, 1117

\bibitem[\protect\citeauthoryear{Foullon et al.}{2011}]{Foullon2011}
    Foullon, C., Verwichte, E., Nakariakov, V.~M., Nykyri, K., Farrugia, C.~J., 2011, ApJ, 729, L8

\bibitem[\protect\citeauthoryear{Foullon et al.}{2013}]{Foullon2013}
    Foullon, C., Verwichte, E., Nykyri, K., Aschwanden, M.~J., Hannah, I.~G., 2013, ApJ, 767, 170

\bibitem[\protect\citeauthoryear{Goedbloed \& Poedts}{2004}]{Goedbloed2004}
    Goedbloed, J.~P., Poedts, S., 2004, \emph{Principles of Magnetohydrodynamics: With Applications to Laboratory and Astrophysical Plasmas}, Cambridge University Press, Cambridge, Chaps.~7 and 11

\bibitem[\protect\citeauthoryear{Hong et al.}{2013}]{Hong2013}
    Hong, J.-C., Jiang, Y.-C., Yang, J.-Y., Zheng, R.-S., Bi, Y., Li, H.-D., Yang, B., Yang, D, 2013, Res.\ Astron.\ Astrophys., 13, 253

\bibitem[\protect\citeauthoryear{Kaiser et al.}{2008}]{Kaiser2008}
    Kaiser, M.~L., Kucera, T.~A., Davila, J.~M., Cyr, O.~C.~St., Guhathakurta, M., 2008, Space Sci.\ Rev., 136, 5

\bibitem[\protect\citeauthoryear{Kamio et al.}{2010}]{Kamio2010}
    Kamio,~S., Curdt, W., Teriaca, L., Inhester, B., Solanki, S.~K., 2010, A\&A, 510, L1

\bibitem[\protect\citeauthoryear{Kiss et al.}{2017}]{Kiss2017}
    Kiss, T.~S., Gyenge, N., Erd\'elyi, R., 2017, ApJ, 835, 47

\bibitem[\protect\citeauthoryear{Kiss et al.}{2018}]{Kiss2018}
    Kiss, T.~S., Gyenge, N., Erd\'elyi, R., 2018, Adv.\ Space Res., 61, 611

\bibitem[\protect\citeauthoryear{Lemen et al.}{2012}]{Lemen2012}
    Lemen, J. R., Title, A.~M., Akin, D.~J., Boerner, P.~F., Chou, C., Drake, J.~F., Duncan, D.~W., Edwards, C.~G., Friedlaender, F.~M., Heyman, G.~F., Hurlburt, N.~E., Katz, N.~L., Kushner, G.~D., Levay, M., Lindgren, R.~W., Mathur, D.~P., McFeaters, E.~L., Mitchell, S., Rehse, R.~A., Schrijver, C.~J., Springer, L.~A., Stern, R.~A., Tarbell, T.~D., Wuelser, J.-P., Wolfson, C.~J., Yanari, C., Bookbinder, J.~A., Cheimets, P.~N., Caldwell, D., Deluca, E.~E., Gates, R., Golub, L., Park, S., Podgorski, W.~A., Bush, R. I., Scherrer, P.~H., Gummin, M.~A., Smith, P., Auker, G., Jerram, P., Pool, P., Soufli, R., Windt, D.~L., Beardsley, S., Clapp, M., Lang, J., Waltham, N., 2012, Sol.\ Phys., 275, 17

\bibitem[\protect\citeauthoryear{Levens et al.}{2016}]{Levens2016}
    Levens, P.~J., Schmieder, B., Labrosse, N., L\'opez Ariste, A., 2016, ApJ, 818,31

\bibitem[\protect\citeauthoryear{Liu et al.}{2009}]{Liu2009}
    Liu, W., Berger, T.~E., Title, A.~M., Tarbell, T.~D., 2009, ApJ, 707, L37

\bibitem[\protect\citeauthoryear{Mart\'inez-Sykora et al.}{2013}]{Martinez-Sykora2013}
    Mart\'inez-Sykora, J., De Pontieu, B., Leenaarts, J., Pereira, T.~M.~D., Carlsson, M., Hansteen, V., Stern, J.~V., Tian, H., McIntosh, S.~W., Rouppe van der Voort, L., 2013, ApJ, 771, 66

\bibitem[\protect\citeauthoryear{Moore et al.}{2013}]{Moore2013}
    Moore, R.~L., Sterling, A.~C., Falconer, D.~A., Robe, D., 2013, Apj, 769, 134

\bibitem[\protect\citeauthoryear{Moore et al.}{2015}]{Moore2015}
    Moore, R.~L., Sterling, A.~C., Falconer, D.~A., 2015, Apj, 806, 11

\bibitem[\protect\citeauthoryear{M\"{o}stl et al.}{2013}]{Mostl2013}
    M\"{o}stl, U.~V., Temmer, M., Veronig, A.~M., 2013, ApJ, 766, L12

\bibitem[\protect\citeauthoryear{Nistic\`o et al.}{2009}]{Nistico2009}
    Nistic\`o, G., Bothmer, V., Patsourakos, S., Zimbardo, G., 2009, Sol.\ Phys., 259, 87

\bibitem[\protect\citeauthoryear{Nistic\`o et al.}{2010}]{Nistico2010}
    Nistic\`o, G., Bothmer, V., Patsourakos, S., Zimbardo, G., 2010, Ann.\ Geophys., 28, 687

\bibitem[\protect\citeauthoryear{Nykyri \& Foullon}{2013}]{Nykyri2013}
    Nykyri, K., Foullon, C., 2013, Geophys.\ Res. Lett., 40, 4154

\bibitem[\protect\citeauthoryear{Ofman \& Thompson}{2011}]{Ofman2011}
    Ofman, L., Thompson, B.~J., 2011, ApJ, 734, L11

\bibitem[\protect\citeauthoryear{Pesnell et al.}{2012}]{Pesnell2012}
    Pesnell, W.~D., Thompson, B.~J., Chamberlin, P.~C., 2012, Sol.\ Phys., 275, 3

\bibitem[\protect\citeauthoryear{Pike \& Mason}{1998}]{Pike1998}
    Pike, C.~D., Mason, H.~E., 1998, Sol.\ Phys., 182, 333

\bibitem[\protect\citeauthoryear{Raouafi et al.}{2016}]{Raouafi2016}
    Raouafi, N.~E., Patsourakos, S., Pariat, E., Young, P.~R., Sterling, A.~C., Savcheva, A., Shimojo, M., Moreno-Insertis, F., DeVore, C.~R., Archontis, V., T\"{o}r\"{o}k, T., Mason, H., Curdt, W., Meyer, K., Dalmasse, K., Matsui, Y., 2016, Space Sci.\ Rev., 201, 1

\bibitem[\protect\citeauthoryear{Shen et al.}{2011}]{Shen2011}
    Shen, Y., Liu, Y., Su, J., Ibrahim,A., ApJ, 735, L43

\bibitem[\protect\citeauthoryear{Su et al.}{2014}]{Su2014}
    Su, Y., G\"{o}m\"{o}ry, P., Veronig, A., Temmer, M., Wang, T., Vanninathan, K., Gan, W., Li, Y., 2014, ApJ, 785, L2

\bibitem[\protect\citeauthoryear{Wedemeyer-B\"{o}hm et al.}{2012}]{Wedemeyer2012}
    Wedemeyer-B\"{o}hm, S., Scullion, E., Steiner, O., Rouppe van der Voort, L., de la Cruz Rodriguez, J., Fedun, V., Erd\'elyi, R., 2012, Nature, 486, 505

\bibitem[\protect\citeauthoryear{Wedemeyer et al.}{2013a}]{Wedemeyer2013a}
    Wedemeyer, S, Scullion, E., Steiner, O., de la Cruz Rodriguez, J., Rouppe van der Voort, L., 2013, J.\ Phys.\: Conference Series, 440, 012005

\bibitem[\protect\citeauthoryear{Wedemeyer et al.}{2013b}]{Wedemeyer2013b}
    Wedemeyer, S, Scullion, E., Rouppe van der Voort, L., Bosnjak, A., Antolin, P., 2013, ApJ, 774, 123

\bibitem[\protect\citeauthoryear{Wedemeyer \& Steiner}{2014}]{Wedemeyer2014}
    Wedemeyer, S., Steiner, O., 2014, PASJ, 66, S10

\bibitem[\protect\citeauthoryear{Young \& Muglach}{2014a}]{Young2014a}
    Young, P.~R., Muglach, K., 2014a, PASJ, 66, S12

\bibitem[\protect\citeauthoryear{Young \& Muglach}{2014b}]{Young2014b}
    Young, P.~R., Muglach, K., 2014b, Sol.\ Phys., 289, 3313

\bibitem[\protect\citeauthoryear{Zank \& Matthaeus}{1993}]{Zank1993}
    Zank, G.~P., Matthaeus, W.~H., 1993, Phys.\ Plasmas, 5, 257

\bibitem[\protect\citeauthoryear{Zaqarashvili et al.}{2015}]{Zaqarashvili2015}
    Zaqarashvili, T.~V., Zhelyazkov, I., Ofman, L., 2015, ApJ, 813, 123

\bibitem[\protect\citeauthoryear{Zhang \& Ji}{2014}]{Zhang2014}
    Zhang, Q.~M., Ji, H.~S., 2014, A\&A, 561, A134

\bibitem[\protect\citeauthoryear{Zhelyazkov et al.}{2015}]{Zhelyazkov2015}
    Zhelyazkov, I., Zaqarashvili, T.~V., Chandra, R., 2015, A\&A, 574, A55

\bibitem[\protect\citeauthoryear{Zhelyazkov et al.}{2015}]{Zhelyazkov2016}
    Zhelyazkov, I., Chandra, R., Srivastava, A.~K., 2016, Astrophys.\ Space Sci., 361, 51

\bibitem[\protect\citeauthoryear{Zhelyazkov et al.}{2018}]{Zhelyazkov2018}
    Zhelyazkov, I., Zaqarashvili, T.~V., Ofman, L, Chandra, R., 2018, Adv.\ Space Res., 61, 628






\end{thebibliography}








\bsp	
\label{lastpage}
\end{document}